{{\PassOptionsToPackage
{spanish,english} {babel}
}}

\documentclass{aa}  
\usepackage{babel}
\usepackage[nottoc]{tocbibind} 
\usepackage{graphicx}
\usepackage{csvsimple}
\usepackage{txfonts}
\usepackage{lscape}
\usepackage{natbib}
\usepackage[utf8]{inputenc}
\bibpunct{(}{)}{;}{a}{}{,}
\usepackage[colorlinks, citecolor=blue]{hyperref}
\usepackage[toc,page]{appendix}
\usepackage{booktabs,caption}
\usepackage[flushleft]{threeparttable}

\newcommand{\hii}{\relax \ifmmode {\mbox H\,{\scshape ii}}\else
H\,{\scshape ii}\fi}
\newcommand{\mi}{\relax \ifmmode {\mu{\mbox m}}\else $\mu$m\fi}
\newcommand{\ha}{\relax \ifmmode {\mbox H}\alpha\else H$\alpha$\fi}
\newcommand{\hb}{\relax \ifmmode {\mbox H}\beta\else H$\beta$\fi}

\newcommand{\sii}{\relax \ifmmode {\mbox S\,{\scshape ii}}\else
S\,{\scshape ii}\fi}
\newcommand{\siii}{\relax \ifmmode {\mbox S\,{\scshape iii}}\else
S\,{\scshape iii}\fi}
\newcommand{\nii}{\relax \ifmmode {\mbox N\,{\scshape ii}}\else
N\,{\scshape ii}\fi}
\newcommand{\oi}{\relax \ifmmode {\mbox O\,{\scshape i}}\else
O\,{\scshape i}\fi}
\newcommand{\oii}{\relax \ifmmode {\mbox O\,{\scshape ii}}\else
O\,{\scshape ii}\fi}
\newcommand{\oiii}{\relax \ifmmode {\mbox O\,{\scshape iii}}\else
O\,{\scshape iii}\fi}
\newcommand{\neiii}{\relax \ifmmode {\mbox Ne\,{\scshape iii}}\else
Ne\,{\scshape iii}\fi}

\newcommand{\rdostres}{\relax \ifmmode {\,\mbox{R}}_{\rm 23}\else
\,\mbox{R}$_{\rm 23}$\fi}

\begin{document}

\title{Unraveling the Kinematics of IZw18: A Detailed Study of Ionized Gas with MEGARA/GTC}

\author{A. Arroyo-Polonio\inst{1},
        C. Kehrig\inst{1},
        J. Iglesias-P\'{a}ramo\inst{1},\inst{2},
        J.M. V\'{i}lchez\inst{1},
        E. P\'{e}rez-Montero\inst{1},
        S. Duarte Puertas\inst{1},\inst{3},
        J. Gallego\inst{4},
        D. Reverte\inst{5},
        A. Cabrera-Lavers\inst{5}}

\institute{Instituto de Astrof\'{i}sica de Andaluc\'{i}a, CSIC, Apartado de correos 3004, 18080 Granada, Spain 
           \and
           Centro Astron\'{o}mico Hispano en Andaluc\'{i}a, Observatorio de Calar Alto, Sierra de los Filabres, 04550 G\'{e}rgal, Spain
           \and
           Departamento de F\'{i}sica Te\'{o}rica y del Cosmos, Universidad de Granada, Granada, Spain
           \and
           Departamento de F\'{i}sica de la Tierra y Astrof\'{i}sica, Universidad Complutense de Madrid
           \and
           Instituto de Astrof\'{i}sica de Canarias, C/ V\'{i}a L\'{a}ctea s/n, La Laguna, Tenerife, Spain
           }

\date{Received ??; accepted ??}

\abstract
{This study offers an in-depth analysis of the kinematic behavior of ionized gas in IZw18, a galaxy notable for its extremely low metallicity and proximity, utilizing data from MEGARA/GTC. We explored the structure and dynamics of the galaxy through H$\alpha$ line profiles, applying single and double Gaussian component fittings to create detailed maps of luminosity, velocity, and velocity dispersion in the main body (MB) and Halo regions. Additionally, we retrieved integrated spectra from various galactic regions to achieve a greater signal-to-noise (S/N) ratio.
In the MB, a rotational pattern is evident, yet a secondary, more complex kinematic pattern emerges from the double-component fitting, further enriched by the identification of a very broad component. Distinguished by a FWHM of nearly 2000 km/s and a wide spatial extension, this component suggests a high-energy outflow and points towards large-scale, non-localized sources of high kinetic energy. Additionally, the observed significant velocity differences between the narrow and very broad components imply that these gases may occupy distinct spatial regions. This is potentially explained by high-density gas near the origin of kinematic input, acting as a 'wall' that reflects back the momentum of the gas.
Regarding the halos, while the NE Halo exhibits a tranquil state with low velocity dispersions, the SW Halo presents higher velocities and more complex kinematics, indicative of diverse dynamic interactions. The identification of the very broad component across the MB and the high kinematical complexity in all regions of the galaxy points towards a scenario of widespread and subtle turbulent motion. 
This nuanced understanding of the kinematic behavior in IZw18, including the interplay of different gas components and the influence of internal structures, enhances our comprehension of the dynamics in blue compact dwarf galaxies. It may provide critical insights into early galaxy formation and the intricate kinematics characteristic of such environments.}


\keywords{Kinematics -- Starburst -- Extreme low metallicity}

\titlerunning{Kinematic analysis of IZw18 with MEGARA/GTC}
\authorrunning{A. Arroyo-Polonio et al.}
\maketitle
\section{Introduction}


The study of galaxy kinematics provides an invaluable tool for understanding the processes that shape the evolution of galaxies. Specifically, analyzing the motion of ionized gas within these celestial structures uncovers patterns indicative of both their current dynamical state and historical phenomena, such as possible rotations and feedback processes.

This study is focused on I Zwicky 18 (hereafter IZw18), which is identified as a blue compact dwarf (BCD) galaxy \citep[e.g.,][]{petrosian1996interferometric,tikhonov2007red,ramos2011spatial} located approximately 18 Mpc away \citep{2017A&A...602A..45L}. IZw18 has attracted attention due to its extremely low metallicity in its ionized gas (12+log(O/H)=7.11 approx 3\% solar; e.g., \cite{kehrig2016spatially}), which places IZw18 among the most metal-poor star-forming systems in the local Universe. Such low metallicity suggests the presence of chemically unevolved stellar populations and makes IZw18 one of the prime candidates for  investigating galaxy formation and evolution processes happening in the primordial universe.

IZw18 has a highly irregular morphology, a characteristic common among BCDs, which further enhances its scientific interest. While its morphology could be the result of recent starburst activity or interactions with other galaxies, the underlying mechanisms remain a subject of active investigation \citep[e.g.,][]{legrand1997detection,legrand1999star,kehrig2015extended}.

In terms of the kinematics of IZw18, the field has been a subject of diverse investigations that have enriched our understanding yet left many questions open. Early works \citep{petrosian1995high,petrosian1996interferometric} noted velocity dispersion values ($\sim$20 km/s) higher than expected from a self-gravitational model \citep{terlevich2015road}.
Furthermore, it was found potential rotational motion in the neutral hydrogen component, leading to speculative theories about the existence of a dark matter halo around the galaxy \citep{van1998complex}. However, these are far from the only intriguing features observed in IZw18.

A study employing long-slit, echelle spectroscopy, specifically exploring the spatial changes in the H$\alpha$ profile, scrutinized the ionized gas within IZw18 in great detail \citep{martin1996kinematic}. This investigation unveiled evidence for a supergiant shell, manifesting as a bipolar outflow in the south-west and north-east extensions of the galaxy, and expanding at speeds between 30 km/s and 60 km/s. However, a subsequent study by \cite{petrosian1996interferometric} presents a contrasting view. It suggests that the perceived bipolar outflow might not be a singular, coherent structure but rather a conglomerate of clumpy HII regions. These regions, presumably powered by star formation, challenge the earlier interpretation of a supergiant shell, proposing a more fragmented and fractal star-forming environment within the galaxy.

In IZw18, disparities in gas kinematics are evident between H$\alpha$ and HeII spectral lines. Notably, HeII lines show a redshift of about 30 km/s compared to H$\alpha$ in regions with peak HeII emission \citep{vaught2021keck}. This discrepancy, potentially linked to supernova remnants or young stellar clusters, suggests a complex and diverse kinematic structure within the galaxy.

IZw18 has been studied as part of a collective analysis of BCDs, as it exemplifies this class of galaxies, characterized by low metallicity (12+log(O/H) between 7.1 and 8.3), irregular and clumpy morphology, high star formation, compact structure, and significant gas richness \citep[e.g.,][]{thuan1981blue,pilyugin1992evolution,izotov1999heavy}. These features, indicative of a young stellar population and ongoing star formation processes, offer insights into the early stages of galactic evolution in low-mass environments. In addition, these galaxies often exhibit chaotic kinematics in HI, overlaying a disk rotation \citep{lelli2012dynamics}. 
In a broader examination of extremely metal-poor galaxies (XMPs)  (i.e. $Z<Z_\odot / 10$), such as IZw18, recent studies like \citep[e.g.,][]{olmo2017kinematics,isobe2023empress,xu2023empress} have observed typical features like rotation in ionized gas, shell outflows, chemical inhomogeneity (although this has not been seen in IZw18 \citep{kehrig2016spatially}) and a high loading factor (over 10), indicative of substantial matter ejection relative to SFR. 
These findings underscore that XMPs are systems influenced by both inflow and outflow mechanisms. 
The extremely low metallicity in such galaxies could be attributed to the accretion of pristine gas, indeed  driven by these inflows. 
Outflows are likely a result of intense stellar feedback, amplified by the low gravitational pull in these low mass galaxies \citep[e.g.,][]{amorin2012complex,bosch2019integral,hogarth2020chemodynamics}. This is evidenced by the broad components in the emission line profiles \citep{olmo2017kinematics}. 

The observational data presented here offer a unique opportunity to deepen our understanding of the ionized gas kinematics of IZw18. The data were obtained using the Multi-Espectr\'{o}grafo en GTC de Alta Resoluci\'{o}n para Astronom\'{i}a (MEGARA) \citep{de2016megara}, an advanced integral field spectrograph capable of capturing detailed spectral data across a two-dimensional field of view (FoV) with high spectral resolution.

By analysing the spectral data of the main body (MB) and two regions in the halo of IZw18, this study seeks to generate a comprehensive kinematic portrait of this intriguing galaxy. This investigation constitutes a step forward in our understanding of the complex kinematics in dwarf galaxies, paving the way for future detailed studies on similar low-mass starburst systems in the Universe.


This paper is organized as follows. In Section \ref{Section:Observations}, we detail the observations and data reduction techniques employed. Section \ref{Section:2D Kinematical analysis} is devoted to the 2D kinematical analysis of the galaxy. The integrated spectra, highlighting spectral features due to enhanced S/N ratio, are presented in Section \ref{Section:Integrated spectra}. In Section \ref{Section:Luminosity weighted velocity}, we discuss the luminosity weighted velocity and velocity dispersion analysis. Finally, Section \ref{Section:Conclusions} summarizes the main conclusions derived from this work. 

\section{Observations}
\label{Section:Observations}

The observational data for this study were acquired using MEGARA, situated at the Folded Cass F focus of the 10.4 m Gran Telescopio Canarias (GTC) telescope at La Palma Observatory \citep{de2016megara}. The Integral Field Unit (IFU) of MEGARA, also known as the Large Compact Bundle (LCB), offers a FoV of 12.5 x 11.3 arcseconds, consisting of 567 hexagonal spaxels, each measuring 0.62 arcseconds. What sets MEGARA apart as a uniquely suitable instrument for this research is its capability to provide two-dimensional spectroscopic data with high spectral resolution. This enables unprecedented detailed two-dimensional kinematic studies of the ionized gas in IZw18, a level of analysis that has not been conducted before. Consequently, the configuration and capabilities of MEGARA make it an invaluable tool for in-depth examinations of individual regions within nearby galaxies.

\begin{figure*}[h!]
\centering
  \resizebox{\hsize}{!}{\includegraphics{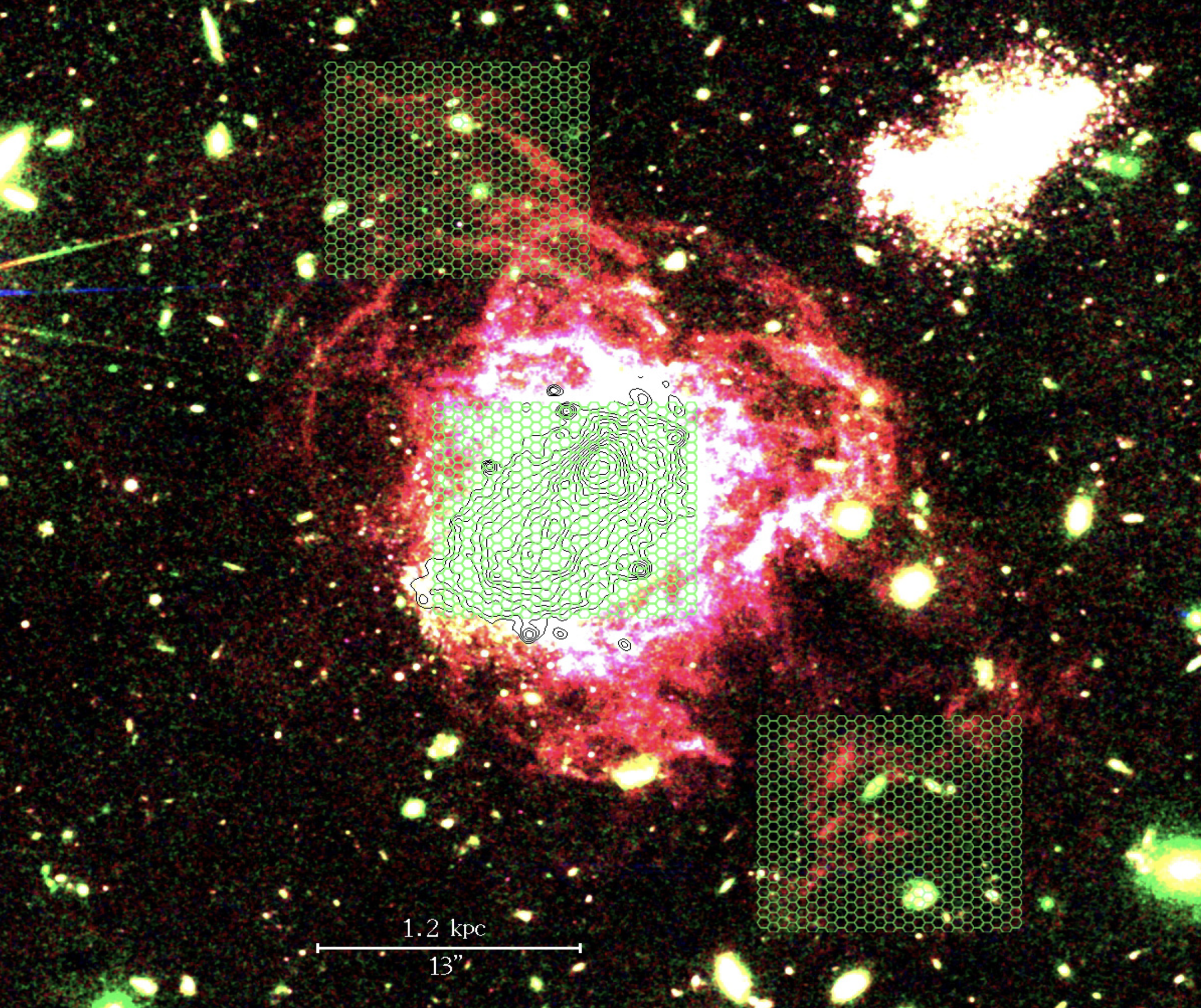}}
  \caption{High contrast composite image of IZw18 using HST ACS R, I, and V bands shown in the red, green, and blue channels, respectively; isocontours delineate the central knots (adapted from \cite{papaderos2012zw}). The three LCB pointings are shown. North is up, East is left here and in all subsequent figures.
}
  \label{apuntado}
\end{figure*}

Our observations focused on three distinct regions within the galaxy IZw18: the MB and two external regions corresponding to the north-east (NE) and south-west (SW) zones of the halo of galaxy. These three regions are represented in Fig. \ref{apuntado}.


The data reduction process was performed using the MEGARA data reduction package \citep{pascual2019megara}, version 0.23.2. This software pipeline allowed for sky and bias subtraction, flat-field correction, spectra tracing and extraction, correction of fiber and pixel transmission, and wavelength calibration. Regarding the sky-subtraction, we utilized the sky fibers available in the
MEGARA instrument, which are specifically designed for this purpose.
MEGARA uses a set of eight dedicated sky fibers that are strategically
placed at the edges of the full FoV, which spans 3.5 arcminutes. These sky
fibers allow for an accurate measurement of the sky background at a
considerable distance from the central object of interest. As the primary objective was a kinematic analysis, no flux calibration was applied. Thus, through this article, all luminosity values are in arbitrary units.

As for the specific settings of the MEGARA instrument during the observations, the LR-R Volume-Phased Holographic (VPH) was employed. This covered a spectral range of 6300$\AA$ - 7363$\AA$ with a spectral resolution of R $\sim$ 6000. For the H$\alpha$ line at a wavelength of 6578Å (redshifted wavelength), this corresponds to a spectral resolution of approximately 1.1$\AA$. In terms of velocity, this spectral resolution is equivalent to approximately 50 km/s.
In Table \ref{table:galaxy_stuff}, we present the log of observations.

\begin{table*}[h]
\centering
\caption{Log of the observations.}
\label{table:galaxy_stuff}
\begin{tabular}{ccccc} 
\hline \hline
Pointing & Exptime            & R.A. & Dec& Date\\ 
         & (s)                &              (J2000.0)&           (J2000.0)&     \\ 
\hline
IZw18 MB & 3 $\times$ 750     & 09:34:02.06 & 55:14:26.5 & 2022-02-06 \\
IZw18 SW & 3 $\times$ 1800    & 09:34:00.37 & 55:14:08.3 & 2022-02-08 \\
IZw18 NE & 3 $\times$ 1500    & 09:34:02.87 & 55:14:44.4 & 2022-02-07 \\
\hline
\end{tabular}

\begin{tablenotes}
      \small
      \item Column (1): Name of each pointing in the observations of the galaxy. Column (2): Exposition time. First it is shown the number of pointings times the exposition time of each one. Columns (3) and (4): Position of the center of each pointing. Column (5): Date of observation.
\end{tablenotes}

\end{table*}

\section{Main kinematical analysis}
\label{Section:2D Kinematical analysis}

In this section we are presenting the main results of this study, which are the maps of luminosity, velocity and velocity dispersion based on the gaussian fit of the H$\alpha$ profile. The methodology followed to construct these maps is also presented here. 

\subsection{Methodology}

The data analysis for this study was executed on a total of six data cubes. These were categorized into two sets: the first set comprised three data cubes (one per pointing) with sky subtraction, while the
second set correspond to the same data cubes but for which no sky
subtraction was performed. 


In the initial phase of our analysis we observed that the center of the [OI]$\lambda$6300 sky line presented spaxel to spaxel variations (up to 0.7 $\AA$). This lead us to hypothesize that the wavelength calibration might not be entirely accurate, at least not to the precision we require for our kinematical analysis. Consequently, we decided to use the [OI]$\lambda$6300 sky line as a reference point for calibrating the velocity of the H$\alpha$ line. For now, to determine the observed central wavelength of each line, we are employing a simple Gaussian fit to each spaxel. 

%

Thus, we defined a reference wavelength field ($\lambda_{ref}$), where the center of the [OI]$\lambda$6300 sky line in each spaxel was taken as the reference wavelength. This $\lambda_{ref}$ was subtracted from the observed H$\alpha$ wavelength (in all the data cubes) field to compute the intrinsic wavelength field of the H$\alpha$ line.
 Importantly, using this methodology, we ensured that any potential inaccuracies in the wavelength calibration were minimized. By referring the position of the H$\alpha$ line to the sky line of [OI]$\lambda$6300, we generated a more accurate wavelength field in H$\alpha$. This effectively neutralized any potential systemic shifts introduced by wavelength calibration uncertainties.

\begin{figure*}[h!]
\centering
  \resizebox{\hsize}{!}{\includegraphics{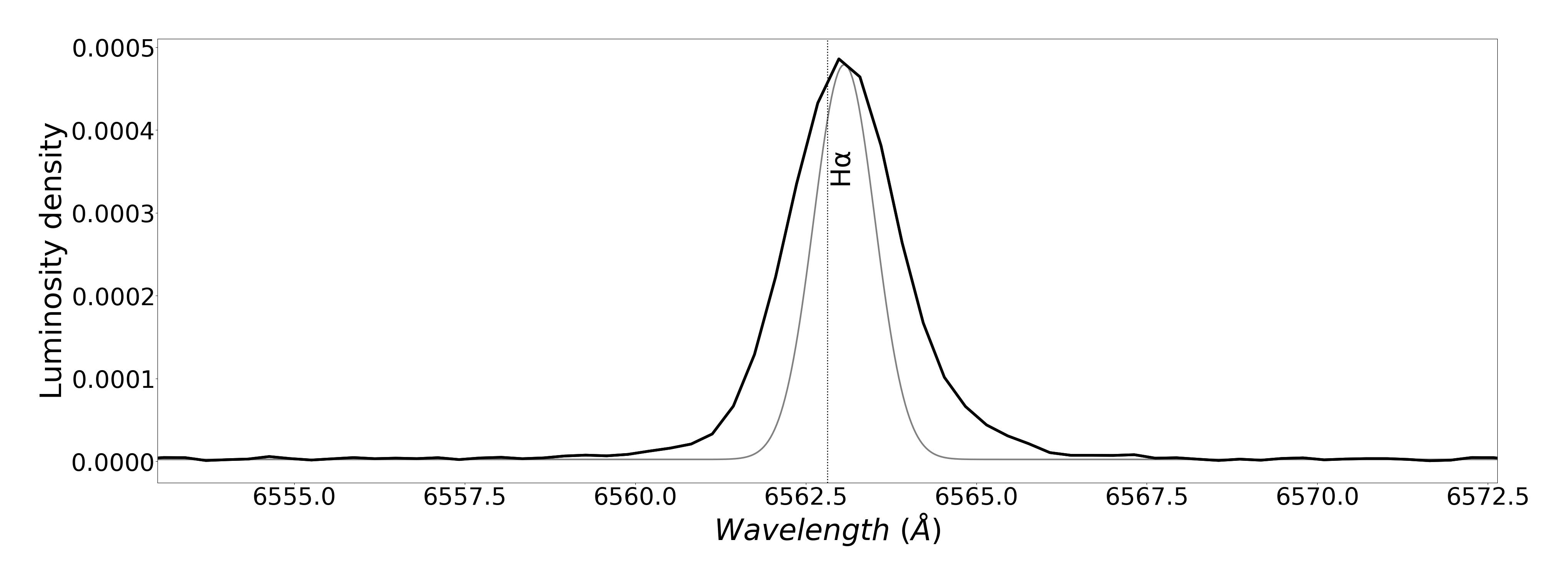}}
  \caption{Black line represents the H$\alpha$ profile of a single spaxel. The gray line is a Gaussian with the instrumental FWHM. The vertical line highlights the H$\alpha$ wavelength.} 
  \label{doble_pico}
\end{figure*}

At this point, attention was shifted to the cubes with sky subtraction. The H$\alpha$ line profiles were examined, and Gaussian fits were applied. Initially, we started with a single Gaussian fit to the H$\alpha$ line profile in each spaxel. However, the observation of broader and more complex profiles in some spaxels led us to explore dual-component fitting. The H$\alpha$ profile of one of these spaxels exhibits a distinctively broadened base with a wavelength shift (see Fig. \ref{doble_pico}), indicative of complex underlying kinematic processes. 

In this way, for each spaxel, both one- and two-component fits were performed, yielding line luminosity, velocity, and velocity dispersion from each Gaussian. Luminosity was calculated from the Gaussian area.
To convert the wavelength field to a velocity field the relativistic Doppler effect was used:

\begin{equation}
v = c \left(\frac{(\lambda/\lambda_0)^2 - 1}{(\lambda/\lambda_0)^2 + 1}\right)
\end{equation}

where $v$ is the velocity, $c$ is the speed of light, $\lambda$ is the observed wavelength, and $\lambda_0$ is the rest wavelength. This Doppler effect conversion was consistently applied in all subsequent velocity calculations.

In order to obtain the corrected velocity dispersion ($\sigma$), we took into account the instrumental and thermal widths ($\sigma_{\mathrm{inst}}$, $\sigma_{\mathrm{thermal}}$). The correction was performed as follows:
\begin{equation}
\sigma^2 = \sigma_{\mathrm{obs}}^2 - (\sigma_{\mathrm{inst}}^2 + \sigma_{\mathrm{thermal}}^2)
\end{equation}
The instrumental width was calculated from the width of the [OI]$\lambda$6300 sky line ($\sigma_{\mathrm{inst}}$=0.45 $\AA$). Meanwhile, $\sigma_{\mathrm{thermal}}$ was calculated by using the Maxwell-Boltzmann distribution:
\begin{equation}
\sigma_{\mathrm{thermal}}^2 = \frac{K T_e}{m}
\end{equation}
where $K$ is the Boltzmann constant, $T_e$ the electronic temperature retrieved from \cite{kehrig2016spatially} (with a value of 20600 K) and $m$ is the hydrogen mass.

In addition to instrumental and thermal widths, we could consider the effects of fine structure and natural broadening ($\sigma_{\mathrm{fs}}$ and $\sigma_{\mathrm{natural}}$). Fine structure, leads to a subdivision of emission lines that can be interpreted as a broadening. The $\sigma_{\mathrm{fs}}$ is only well characterised for atoms with one electron and is usually around 2-3 km/s \citep[e.g.,][]{garcia2008velocity} so for our spectral resolution we can ignore this effect. Natural broadening, governed by the Heisenberg uncertainty principle ($\Delta E \cdot \Delta t \geq \hbar/2$) in our case is minimal. Lifetimes of excited states of the H atom are about 10 nanoseconds, leading to a natural linewidth broadening of only a few m/s, which is significantly smaller than other broadening mechanisms encountered in astrophysical observations.


In our one-component fit analysis, certain spaxels are excluded to maintain data integrity. This exclusion is based on three criteria: spaxels where the FWHM of the H$\alpha$ line is narrower than the instrumental FWHM, discrepancies between the H$\alpha$ and H$\beta$ line profiles, and a Signal/Noise ratio below 3, where Signal represents the peak flux of the H$\alpha$ line and Noise denotes the standard deviation of the continuum beneath the line. Those spaxels were tacked away from the representation of the luminosity, velocity and the velocity dispersion maps.
In the two-component fitting process, we applied two extra conditions. One crucial condition was using the Akaike Information Criterion (AIC). If any H$\alpha$ profile (at a particular spaxel) was better fitted by the single-component fit as indicated by a lower AIC value, it was excluded from the dual-component analysis. Additionally, any Gaussian with an amplitude less than 2.5 times the standard deviation of the continuum was disregarded in the two-component fit.

Due to the high precision required for the two-component fits, all the data cubes were spatially smoothed prior to the fitting process using a Gaussian kernel with a FWHM of 1.4 ''. This specific choice of FWHM was arrived after exploring several kernel sizes, aiming to optimize the trade-off between enhancing coherence among adjacent spaxels and minimizing the loss of spatial resolution. The smoothing process, therefore, not only improved the coherence between adjacent spaxels but did so while carefully considering the inevitable compromise in spatial resolution. In \cite{ebeling2006asmooth} it is shown the high utility of using gaussian kernels for smoothing astronomical images.

Finally, it was necessary to distinguish between the primary and secondary components of each dual-component fit. This distinction was made using two criteria: the residual area after subtracting the two Gaussians of the dual fit from the single-component Gaussian, and the absolute velocity difference between each of the dual component velocities and the single component velocity.

For the velocity difference criterion, we defined the velocity of the single-component fit as $v_{\mathrm{single}}$, and the velocities of the two components of the dual fit as $v_{1,\mathrm{dual}}$ and $v_{2,\mathrm{dual}}$. Then, we computed the absolute velocity differences, $|v_{\mathrm{single}} - v_{1,\mathrm{dual}}|$ and $|v_{\mathrm{single}} - v_{2,\mathrm{dual}}|$. The component from the dual fit with the smallest absolute velocity difference was identified as the primary component, thus ensuring that the primary component velocity was as similar as possible to the single component velocity.

For the residual area criterion, we compared the residual areas, where $G_{\mathrm{single}}(\lambda)$ represents the Gaussian fit for the single-component model, and $G_{1,\mathrm{dual}}(\lambda)$ and $G_{2,\mathrm{dual}}(\lambda)$ are the Gaussian fits for the first and second components of the dual fit, respectively. These residual areas are defined as $A_{\mathrm{residual_1}} = \int |G_{\mathrm{single}}(\lambda) - G_{1,\mathrm{dual}}(\lambda)| d\lambda$ and $A_{\mathrm{residual_2}} = \int |G_{\mathrm{single}}(\lambda) - G_{2,\mathrm{dual}}(\lambda)| d\lambda$. The component with the lower residual area, more closely aligning with the single-component fit, was designated as the primary component.

As a final remark, every calculation of the velocity is relative to the velocity of the center of mass of the ionized gas emitting H$\alpha$. Since the mass of gas is proportional to the luminosity, we can set the velocity of the center of mass as the zero velocity.

\subsection{Luminosity, velocity and velocity dispersion maps}
\label{subsection: Luminosity, velocity and velocity dispersion maps}

The result of this methodical process was a comprehensive suite of nine maps including each of the three pointings (MB, NE halo and SW halo). Each one displaying luminosity, velocity, or velocity dispersion of the H$\alpha$ line for the three different components (the two components of the dual-fit and the one in the single-fit).

The maps are designed to offer a holistic yet detailed view of the different regions of the galaxy. Each map features a background layer sourced from deep HST observations (same as used in Fig. \ref{apuntado}), which cover a wide field of view to capture the galaxy in its entirety. This panoramic backdrop serves to contextualize the more targeted data overlays from the MEGARA/GTC observations. Specifically, three distinct overlays are superimposed on the HST background in each map, each corresponding to one of the observed regions of the galaxy: the MB, the SW halo, and the NE halo.



 \subsubsection{Single-componnent}

\begin{figure*}[h!]
\centering
  \resizebox{\hsize}{!}{\includegraphics{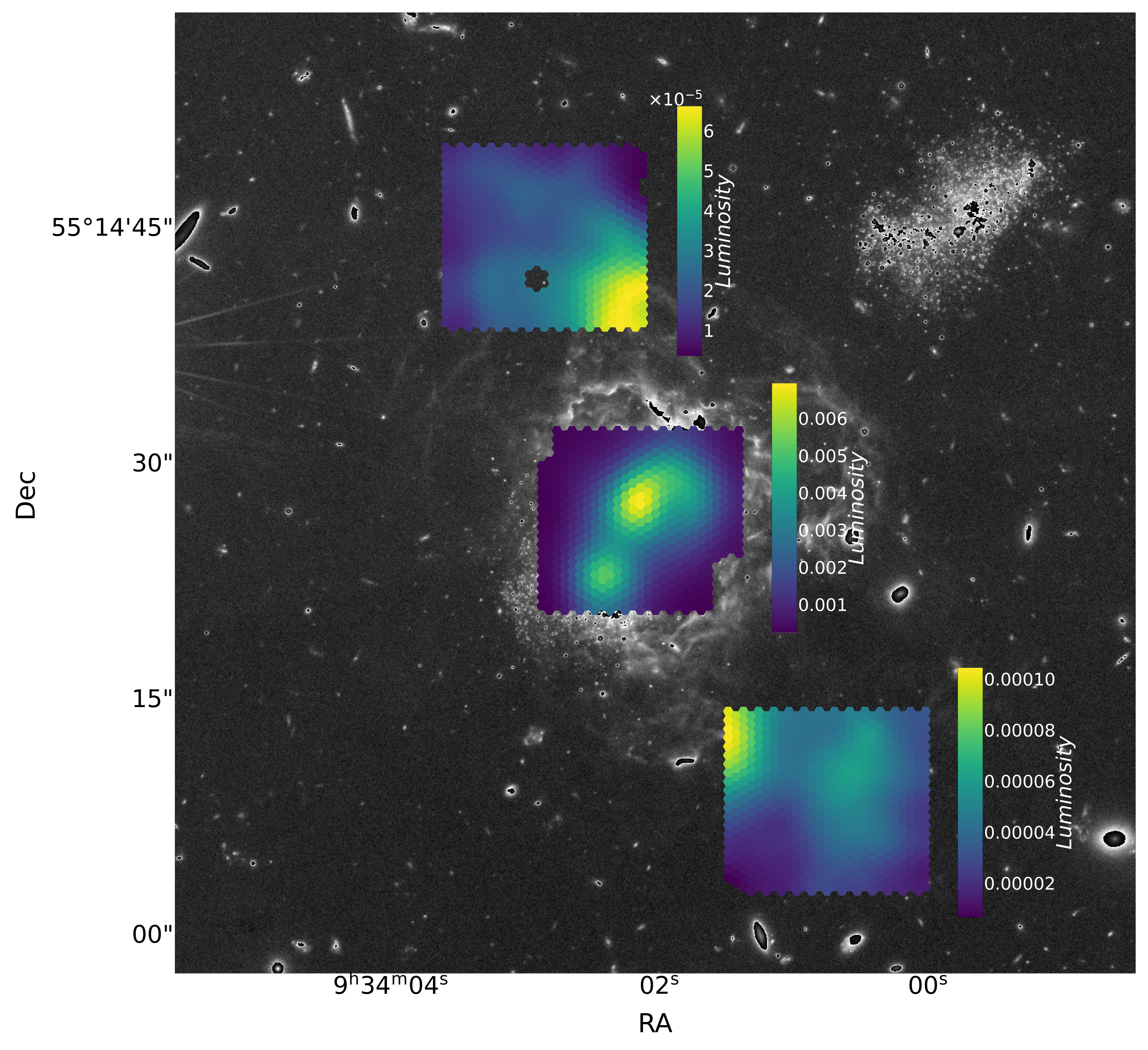}}
  \caption{H$\alpha$ luminosity maps for the MB and Halos obtained from the single-component fit. The background layer, presented in a cyclical gray scale, represents the wide-field deep image from the HST using the f814w and f606w filters of the WFC2 instrument. This background is used in all the images of IZw18 where the MB and both Halos are represented.}
  \label{Luminosidad_1_Gauss}
\end{figure*}

We begin our exploration with the maps derived from a single-component Gaussian fit.
Initially, attention is drawn to the luminosity map, depicted in Fig. \ref{Luminosidad_1_Gauss}, which delineates the distribution of H$\alpha$ luminosity across the different regions of the galaxy IZw18. The H$\alpha$ luminosity distribution serves as a bedrock for comprehending the structure of the ionized gas within the galaxy.
The luminosity
map of the Izw18 MB distinguishes three main features which are pretty well
known: two star-forming knots, one situated south-east and the other north-west, and arc-like structure (‘plume’) which has one end rooted in the vicinity
of the NW knot \citep[e.g.,][]{davidson1989high,dufour1990extended,dufour1996evidence,vilchez1998bidimensional,papaderos2012zw,kehrig2016spatially}. These features underscore the known morphology of IZw18, presenting a lucid portrayal of the distribution and nature of ionized gas, which is particularly insightful for the kinematic study being undertaken. When focusing on the two halos, it is apparent from the luminosity map that their luminosity is significantly diminished compared to that of the MB (around 1 or 2 orders of magnitude). Moreover, these maps also trace the outer shells of IZw18, as we can see comparing with the background HST image, thereby providing a nuanced understanding of the ionized gas distribution and its overarching structure.

\begin{figure*}[h!]
\centering
  \resizebox{\hsize}{!}{\includegraphics{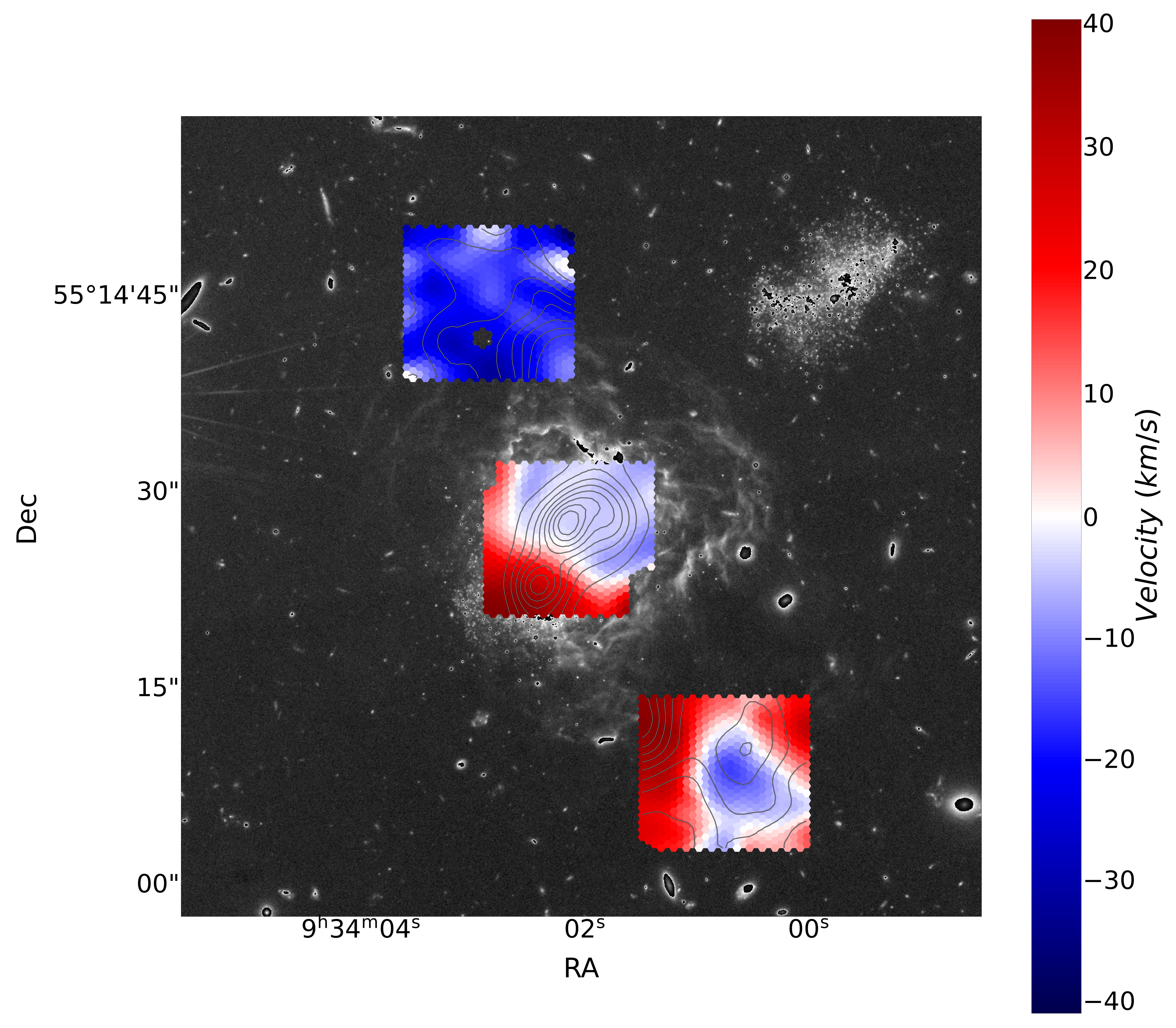}}
  \caption{H$\alpha$ velocity maps for the MB and Halos obtained from the single-component fit. The contours correspond to the H$\alpha$ luminosity in log scale obtained from the same fit component. The same contours are represented in all H$\alpha$ velocity and velocity dispersion maps.}
  \label{Velocidad_1_Gauss}
\end{figure*}

Following the examination of the luminosity map, the focus transitions to the velocity map illustrated in Fig. \ref{Velocidad_1_Gauss}. This map emerges as an indispensable tool for unraveling the kinematic behavior of the ionized gas within IZw18, offering insights into the internal motion within both the MB and the halos, and unearthing patterns and flows that are pivotal to understand the dynamical state of the galaxy.

Within the MB, the velocity map distinctly unveils a rotation pattern between the two central knots, with a velocity difference of $\sim$ 40 km/s. This harmonize with findings from several studies that have explored such kinematics in H$\alpha$ and HI using the 21 cm line \citep{petrosian1996interferometric,van1998complex,lelli2012dynamics,vaught2021keck,isobe2023empress,xu2023empress}. Notably, this rotational feature seems to dissipate towards the north-east and south-west peaks of the MB map, where a shift towards redder velocities (i.e., positive velocities or gas moving away from us) is observed, indicative of outward or recessionary motion.

Transitioning to the halos, the velocity maps presented herein are pioneering, unveiling for the first time the 2D kinematics of ionized gas in these outer realms of IZw18 and tracing the movement of the shells enveloping the galaxy. Previous studies \citep[e.g.,][]{vaught2021keck} studied the kinematics of the H$\alpha$ line up to $\sim$ 0.75 kpc from the center of the galaxy. In our study we observe these two regions of the halo up to $\sim$ 3 kpc away from the center. Particularly in the NE halo, a relatively uniform map is observed, with velocities leaning towards the blue albeit with some velocity gradients present. Conversely, the SW halo exhibits a more complex kinematic structure, with gas manifesting a large scale turbulent motion, possibly hinting at underlying dynamical processes or interactions occurring within this region.

\begin{figure*}[h!]
\centering
  \resizebox{\hsize}{!}{\includegraphics{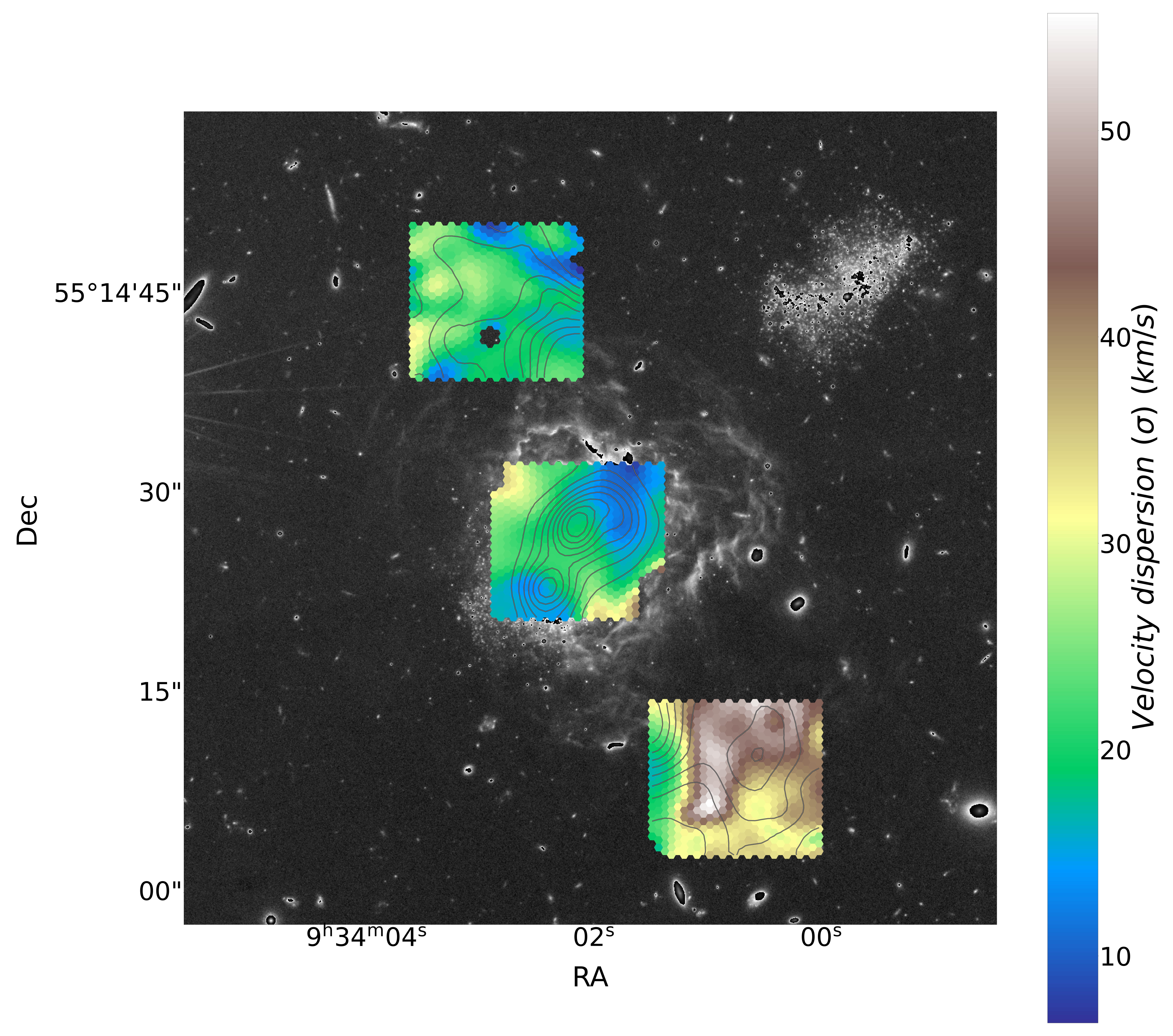}}
  \caption{H$\alpha$ velocity dispersion maps for the MB and Halos obtained from the single-component fit.}
  \label{Dispersion_1_Gauss}
\end{figure*}

Wrapping up our exploration of single-component maps, the velocity dispersion maps depicted in Fig. \ref{Dispersion_1_Gauss} merits a close examination. This map elucidates the spectrum of velocities inherent within each region, serving as a discerning gauge of areas of dynamic stability or potential turbulence.

A foundational aspect to consider is the theoretical velocity dispersion IZw18 would exhibit if the gas in the galaxy adheres to the virial theorem, as delineated in \cite{terlevich2015road}. We retrieved the H$\beta$ integrated luminosity from the H$\beta$ flux ($1.59\times10^{-13} \ erg \ s^{-1} \ cm^{-2}$ ) in \cite{kehrig2016spatially} and the distance (18.2 Mpc) in \cite{aloisi2007zw}. From this H$\beta$ luminosity ($6.3\times10^{39} \ erg \ s^{-1}$) we conclude according to \cite{terlevich2015road} that if the observed velocity dispersion surpasses the threshold of $\sim$ 20 km/s, it signifies the presence of another process (beyond the natural random motion induced by gravitational interaction) that escalate the kinetic energy of the gas, thereby augmenting the velocity dispersion.

In scrutinizing the map, it is notable that a majority of the regions encompassing the MB and the NE halo exhibit velocity dispersions < 25 km/s. This suggests that, broadly, these velocity dispersions could predominantly be attributed to the virial motion of the gas. However, within the MB, particularly in the top left and bottom right corners (see again Fig. \ref{Dispersion_1_Gauss}), velocity dispersions surge to up to 40 km/s. This escalation hints at broadening or added complexity in the H$\alpha$ profile in these regions. Regarding the Halos, the NE Halo presents a pretty mild velocity dispersion which do not go above 35 km/s indicating that there is not much turbulent motion in this region. The SW Halo nevertheless presents the higher value going up to 55 km/s in most regions where the velocity is zero (see regions with white velocity according to the colorbar in Fig. \ref{Velocidad_1_Gauss} and zones with brown - white velocity dispersion in Fig. \ref{Dispersion_1_Gauss}).

\begin{figure*}[h!]
\centering
  \resizebox{\hsize}{!}{\includegraphics{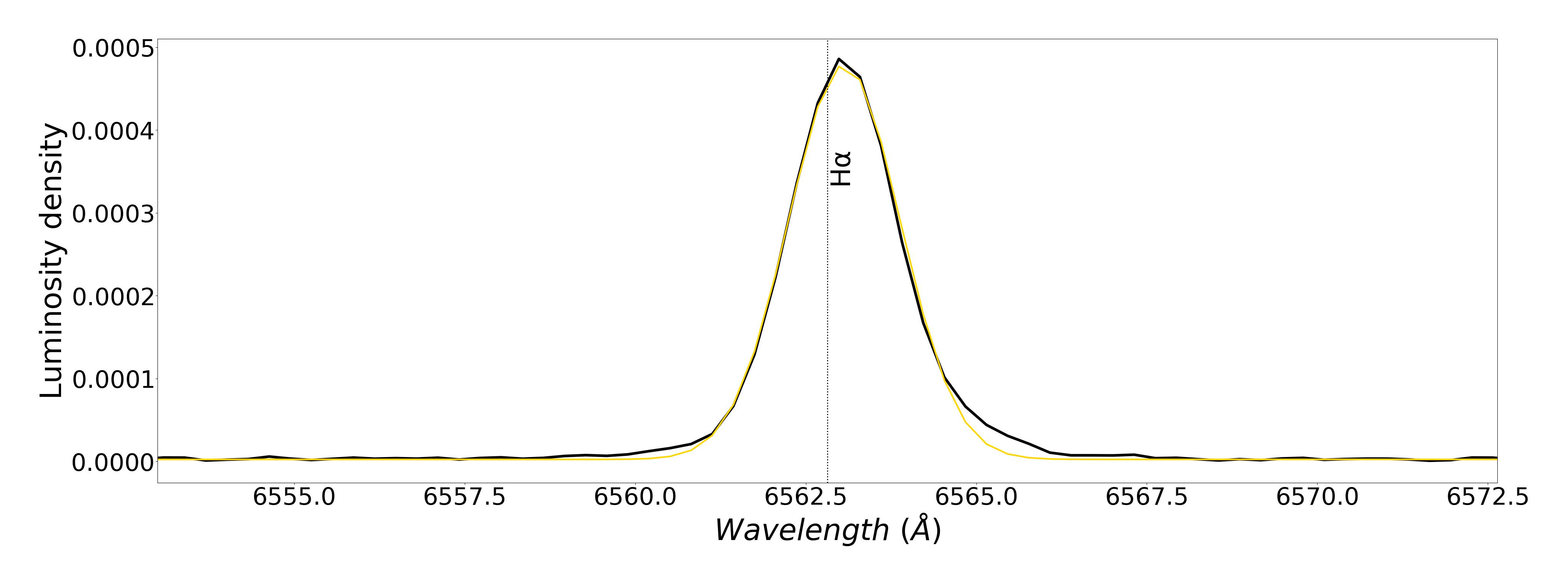}}
  \caption{One-component fit of a H$\alpha$ profile of a single spaxel in the south-west of the MB (same spaxel as in Fig. \ref{doble_pico}) of the galaxy. The profile is depicted in black, meanwhile the one-component fit is represented in yellow.}
  \label{doble_pico_1_gauss}
\end{figure*}

The corresponding one-component fit of the spectrum shown in Fig. \ref{doble_pico} is represented in Fig. \ref{doble_pico_1_gauss}. As we can see, this simple fit does not capture all the complexities inherent in the observed line profile, which exhibits broadening in the base. Regions where the single-component fit shows high velocity dispersions are often indicative of a more complex or broad H$\alpha$ line profile. 

 \subsubsection{Double-componnent}

\begin{figure*}[h!]
\centering
  \resizebox{\hsize}{!}{\includegraphics{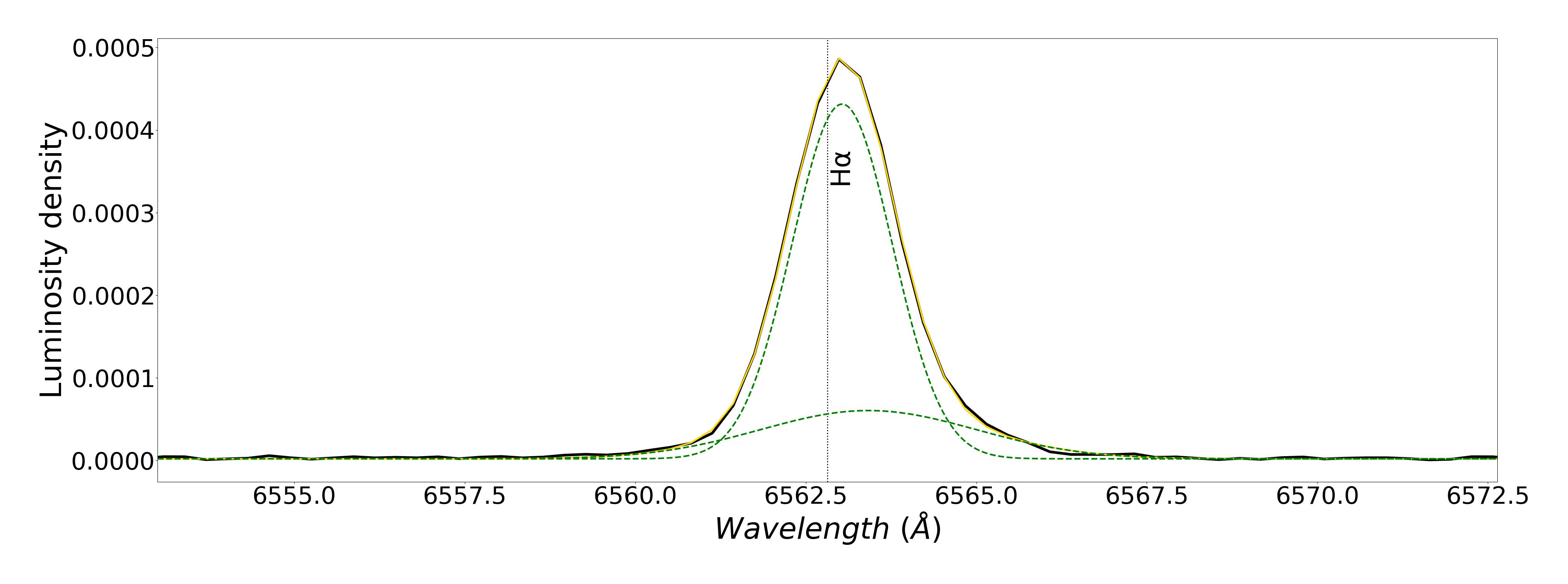}}
  \caption{Double-component fit of a H$\alpha$ profile of a single spaxel in the south-west of the MB (same spaxel as in Fig. \ref{doble_pico}) of the galaxy. The profile is depicted in black, meanwhile the double-component fit is represented in yellow. The green lines represent each of the gaussian corresponding to the double-component fit.}
  \label{doble_pico_s_gauss}
\end{figure*}

In contrast, a double-component Gaussian fit, applied to the same spectrum, provides a more nuanced and accurate representation (see Fig. \ref{doble_pico_s_gauss}). The small, broad Gaussian component successfully fits the broadening in the base of the profile, thereby resolving the complexity in the profile. This dual-component approach proves particularly valuable in regions where the single-component fit is inadequate, as it disentangles the overlapping velocity structures and offers a clearer understanding of the internal dynamics of IZw18.

Expanding our discussion to encompass the double-component Gaussian fits, we simultaneously examine both the principal and secondary components. In doing so, we aim to provide a more comprehensive understanding of the intricate dynamics at play within the different regions of the galaxy.

\begin{figure*}[h!]
\centering
\resizebox{\hsize}{!}{\includegraphics{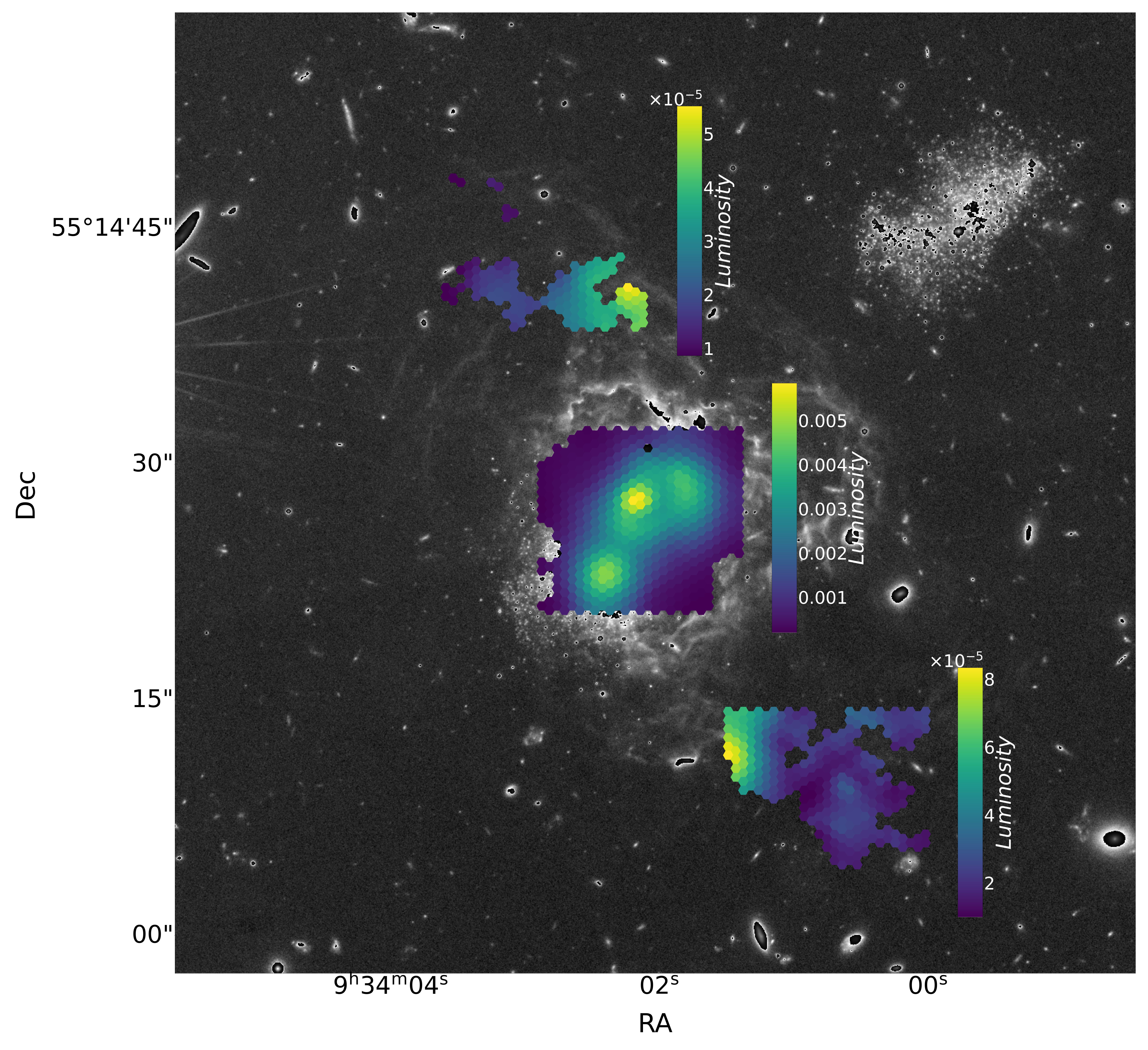}}
\caption{H$\alpha$ luminosity maps for the MB and Halos corresponding to the principal double-component fit.}
\label{Luminosidad_2_Gauss_Principal}
\end{figure*}

\begin{figure*}[h!]
\centering
\resizebox{\hsize}{!}{\includegraphics{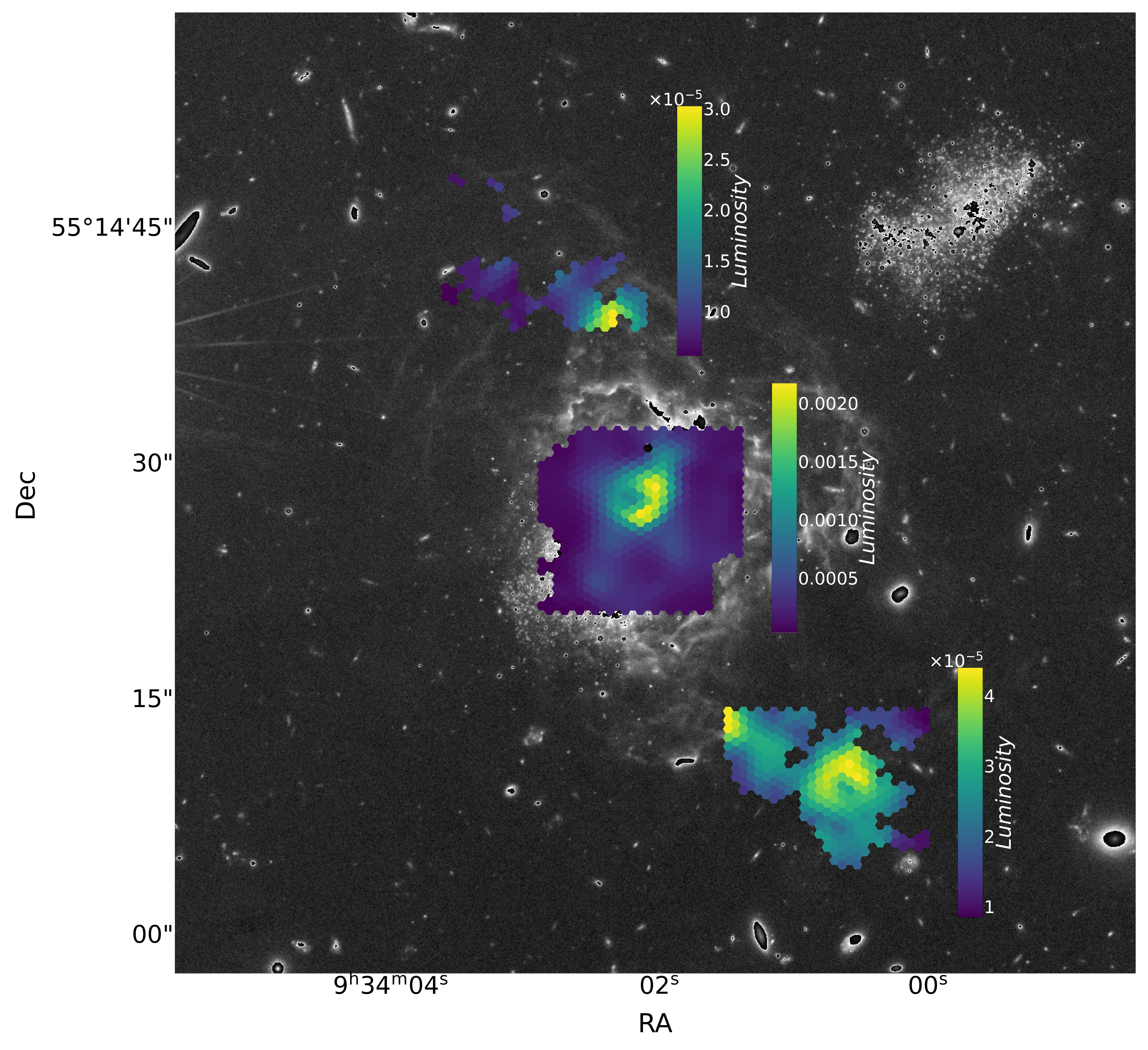}}
\caption{H$\alpha$ luminosity maps for the MB and Halos corresponding to the secondary double-component fit.}
\label{Luminosidad_2_Gauss_Secundaria}
\end{figure*}

We initiate our discussion with the H$\alpha$ luminosity maps, depicted in Figs. \ref{Luminosidad_2_Gauss_Principal} and \ref{Luminosidad_2_Gauss_Secundaria} for the principal and secondary components, respectively. It is noteworthy that the sum of the luminosities from the principal and secondary components in each spaxel closely approximates the luminosity captured by the single-component fits. This indicates a high degree of correspondence between the two approaches, yet with the double-component fit offering a more nuanced view. The principal component is typically more luminous compared to the secondary component (around 4 times more in the MB, 2 times more in the NE Halo and similar in the SW Halo), fitting well with the two knots in the MB of the galaxy. Meanwhile, the secondary component appears to trace the plume of the galaxy.


\begin{figure*}[h!]
\centering
\resizebox{\hsize}{!}{\includegraphics{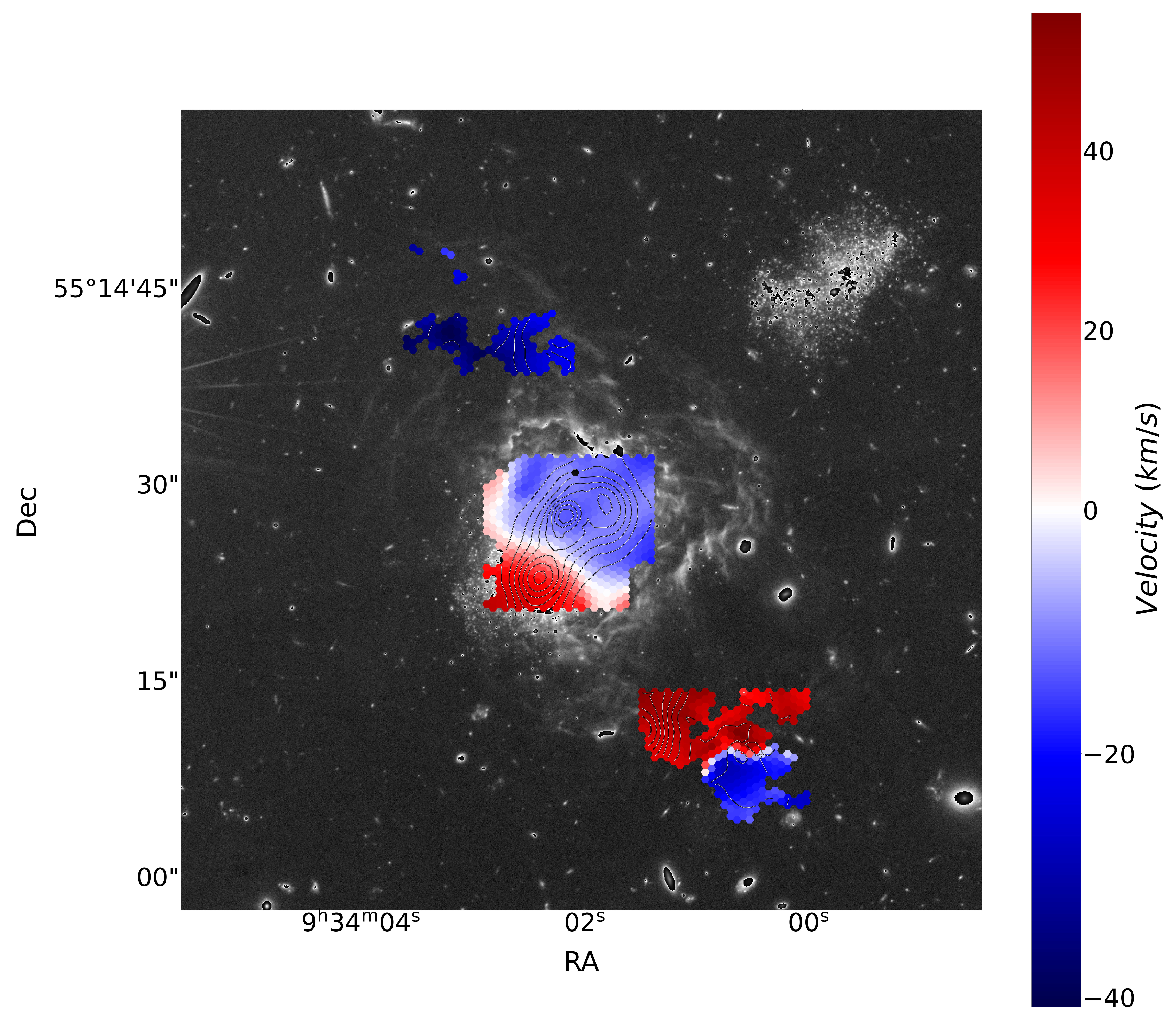}}
\caption{H$\alpha$ velocity maps for the MB and Halos corresponding to the principal double-component fit.}
\label{Velocidad_2_Gauss_Principal}
\end{figure*}

\begin{figure*}[h!]
\centering
\resizebox{\hsize}{!}{\includegraphics{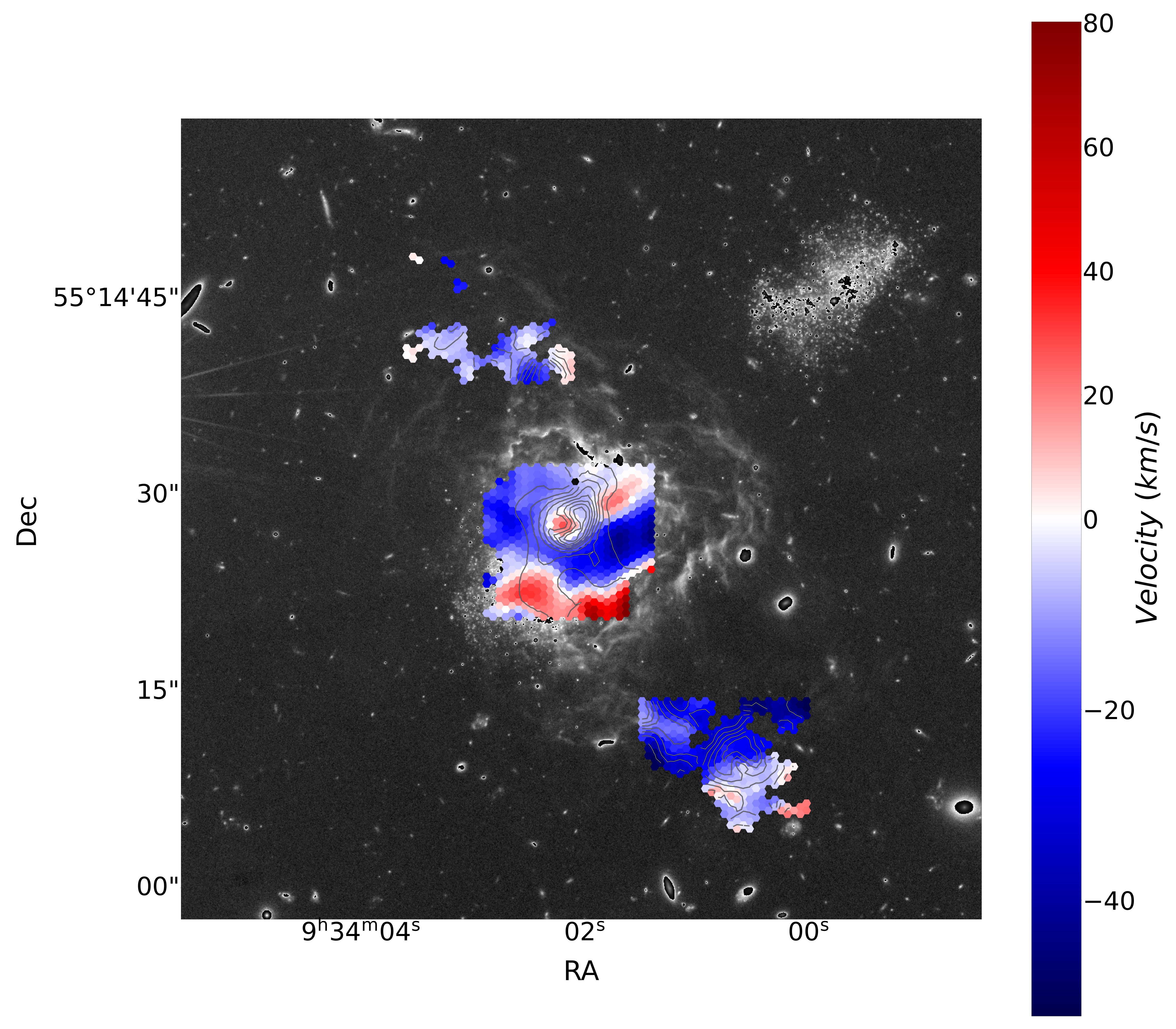}}
\caption{H$\alpha$ velocity maps for the MB and Halos corresponding to the secondary double-component fit.}
\label{Velocidad_2_Gauss_Secundaria}
\end{figure*}

Transitioning to the H$\alpha$ velocity, the maps presented in Figs. \ref{Velocidad_2_Gauss_Principal} and \ref{Velocidad_2_Gauss_Secundaria} elucidate the internal kinematics of the ionized gas in IZw18. Remarkably, the velocity map of the principal component bears a close resemblance to that of the single-component fit, both exhibiting patterns indicative of rotation. This similarity is especially pronounced in regions where the single-component velocity dispersion is low (<25 km/s), suggesting that the H$\alpha$ profile is less complex and that the single-component fit is sufficiently accurate in those areas. 
However, this is not the case for the SW Halo, where the velocity of the principal component divides this region into two distinct zones, north and south, with markedly different velocities. The transition between these zones is abrupt, indicating a significant variation in the kinematic properties of the SW Halo.

The secondary component presents a significantly different picture. Its velocity range is more extensive than that of the principal component (principal component goes from -40 km/s to 55 km/s, secondary instead goes from -55 km/s to 80 km/s), possibly pointing to higher velocities in more subtle areas of the galaxy. The shape and distribution of velocities in this secondary component are also more challenging to interpret. While the principal component largely aligns with previously studied HI gas rotations \citep[e.g.,][]{lelli2012dynamics}, the secondary component could either be tracing different gases in different regions or potentially be indicative of large-scale turbulent motions within the galaxy.


\begin{figure*}[h!]
\centering
\resizebox{\hsize}{!}{\includegraphics{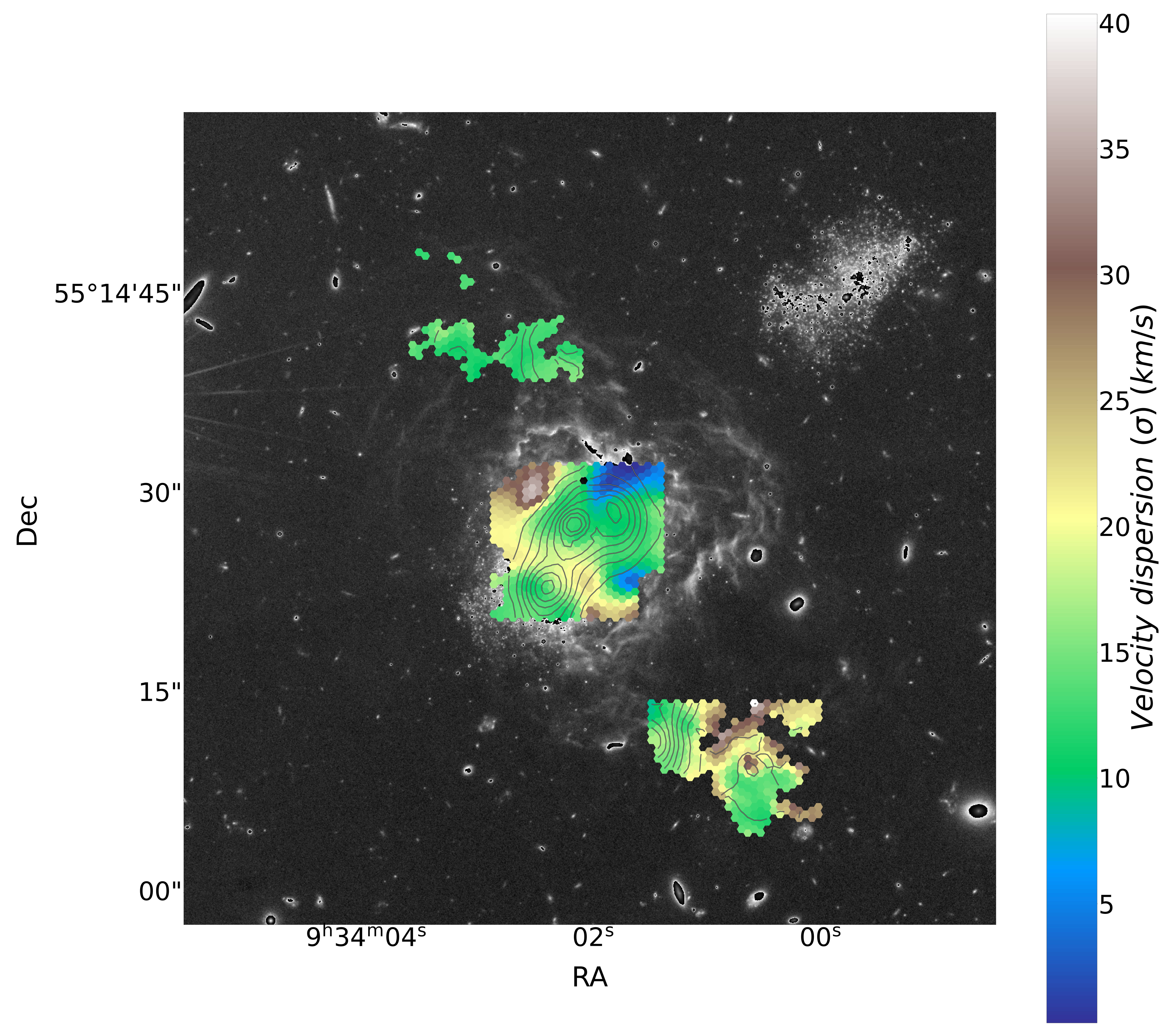}}
\caption{H$\alpha$ velocity dispersion maps for the MB and Halos corresponding to the principal double-component fit.}
\label{Dispersion_2_Gauss_Principal}
\end{figure*}

\begin{figure*}[h!]
\centering
\resizebox{\hsize}{!}{\includegraphics{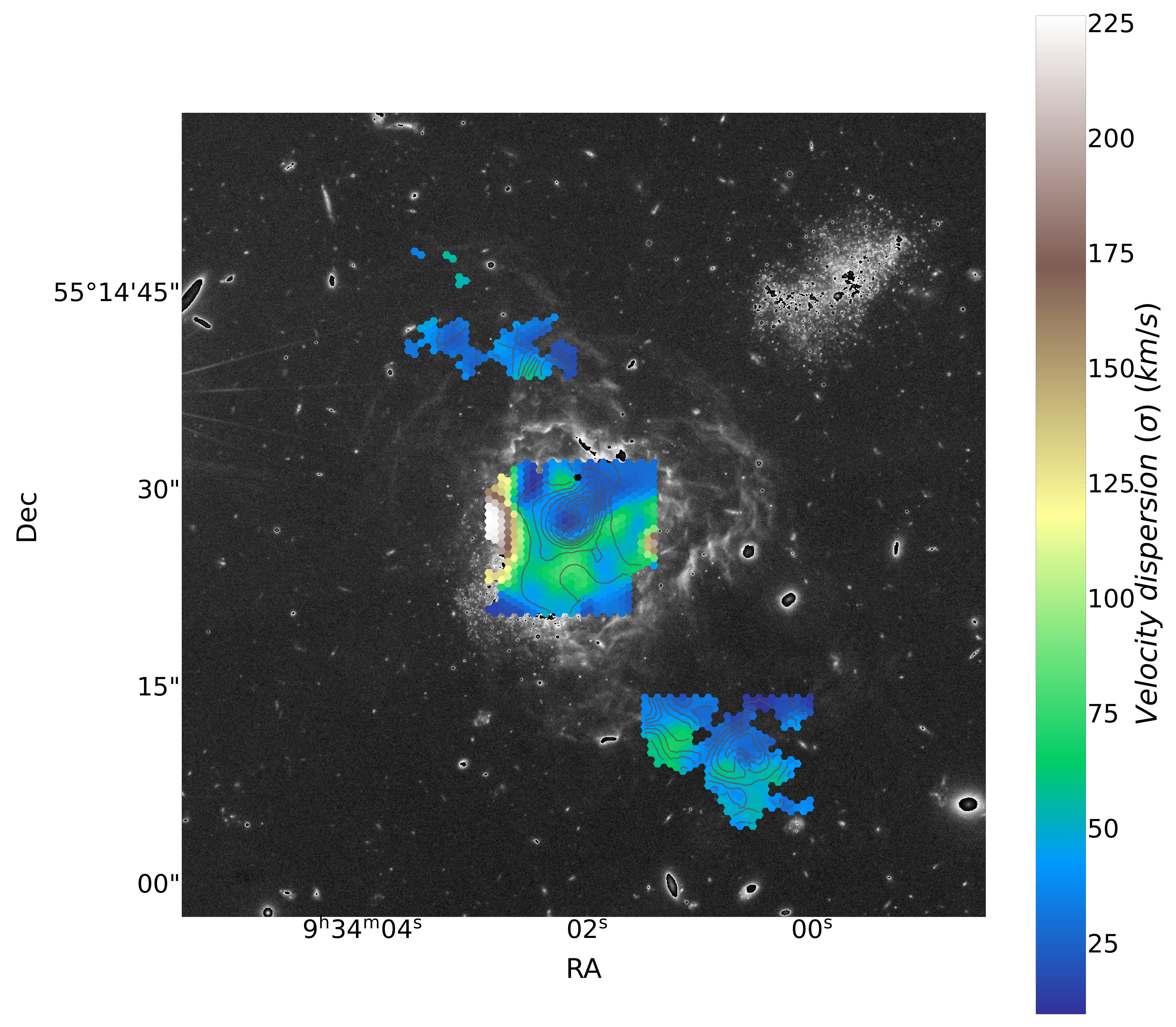}}
\caption{H$\alpha$ velocity dispersion maps for the MB and Halos corresponding to the secondary double-component fit.}
\label{Dispersion_2_Gauss_Secundaria}
\end{figure*}

Lastly, we explore the H$\alpha$ velocity dispersion maps for both components, as shown in Figs. \ref{Dispersion_2_Gauss_Principal} and \ref{Dispersion_2_Gauss_Secundaria}. The velocity dispersion in the principal component closely resembles that of the single-component fit, particularly in regions where both are low (<25 km/s). Notably, in areas where the single-component dispersion is high (>25 km/s), the principal component often shows a lower dispersion value, hinting at the limitations of the single-component approach in capturing the full dynamical complexity.

The secondary component, on the other hand, presents a markedly different scenario. Its velocity dispersion values are up to more than five times greater than those retrieved from either the single or principal components, underscoring the presence of high turbulence and or kinematical complexity in more nuanced regions of the galaxy. Most strikingly, the eastern (and with less relevance also the western) part of the MB exhibits velocity dispersions reaching up to 225 km/s, potentially tracing underlying outflows in this particular zone of IZw18.
A 3D representation of all this kinematical features of the observed regions of IZw18 can be seen in Appendix \ref{appendix: 3D IZw18}.


\subsection{Luminosity weighted velocity}
\label{Section:Luminosity weighted velocity}

In this section, we delve into the analysis of the kinematics of IZw18 through luminosity-weighted histograms for velocity and velocity dispersion, based on the maps discussed in \ref{subsection: Luminosity, velocity and velocity dispersion maps}. The core of this approach lies in weighting the velocity and velocity dispersion values by the luminosity associated with each spaxel, which is derived from the H$\alpha$ emission line. This luminosity is directly proportional to the number of ionized hydrogen atoms, and thus to the mass of ionized gas, providing a mass-weighted perspective of the kinematic properties. By employing this method, we aim to emphasize the contributions from regions with higher mass concentrations, offering a more nuanced understanding of the dynamic behavior of the galaxy.

To construct the velocity histograms, we start by establishing the range of velocities, defined by the maximum and minimum values observed in the velocity map. This range is then divided into uniformly sized bins. For each bin, we aggregate the luminosities of all spaxels whose velocities fall within the range of the bin, effectively creating a sum of luminosities that represent the mass of ionized gas moving at those velocities. Normalizing these sums by the total luminosity of all spaxels yields the luminosity-weighted velocity histogram. A peak in this histogram indicates a predominant velocity at which a significant mass of ionized gas is moving, providing insight into the dominant kinematic flows of the galaxy.

Similarly, the velocity dispersion histograms are constructed using the same luminosity-weighting approach, allowing us to assess the spread of velocities around the mean motion in a mass-weighted manner. This method enriches our understanding of the internal motions within the ionized gas, revealing the complexities of its dynamics and the forces at play.

\begin{figure*}[h!]
\centering
\resizebox{\hsize}{!}{\includegraphics{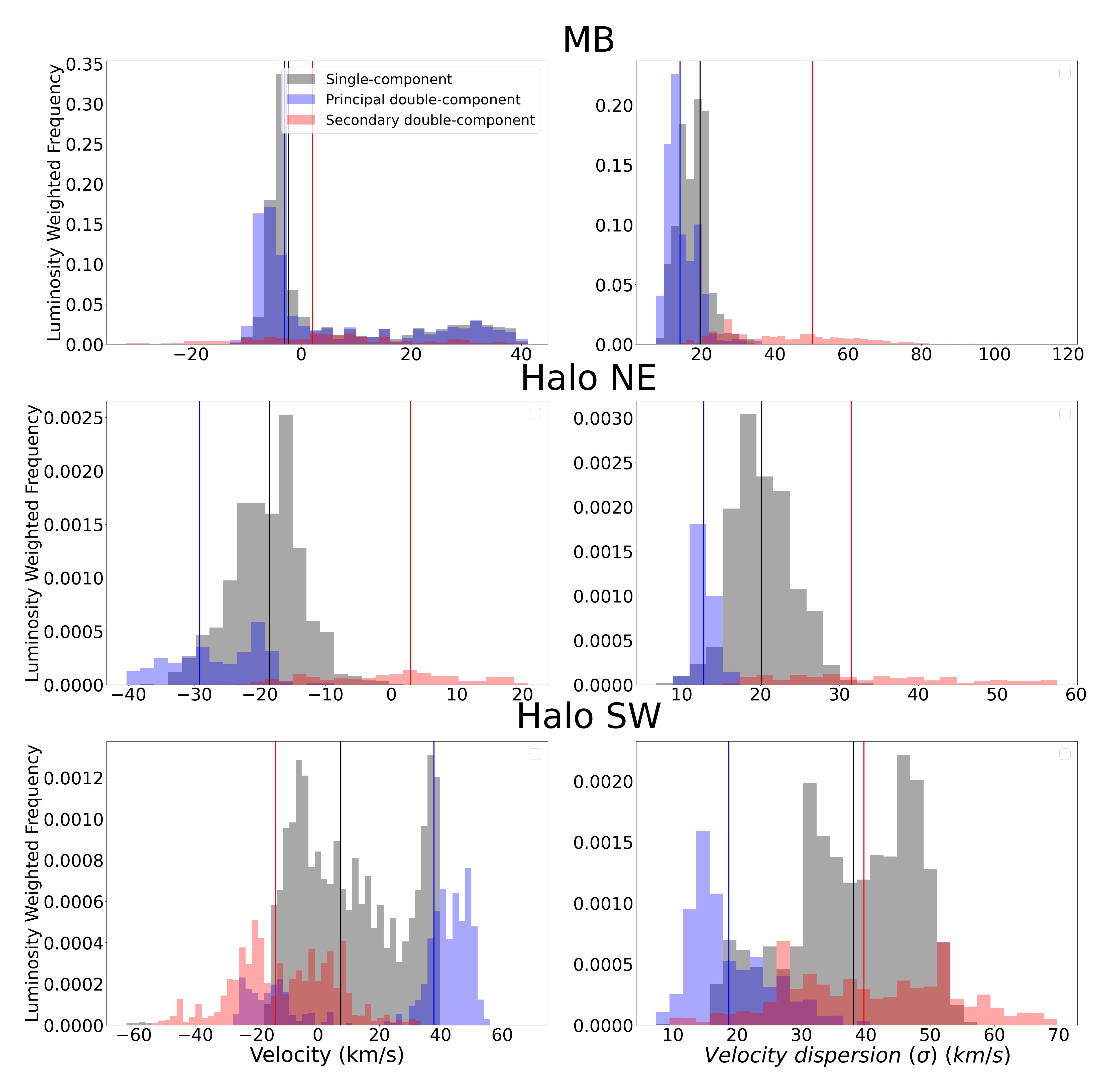}}
\caption{Histograms of the luminosity-weighted velocity and velocity dispersion.
Top row corresponds to the MB, middle row to the halo NE and bottom row to the halo SW. In the left column we represent the luminosity-weighted velocity and in the right column it is represented the luminosity-weighted velocity dispersion.
The black color represent calculations made with the single-component fit, meanwhile blue and red colors represent the principal and secondary double-component fit, respectively.
Vertical solid thick lines represents the median of each distribution.
The width of each cell in the x axis is selected to be 2 km/s}
\label{Histogramas}
\end{figure*}

Starting with the MB of IZw18 in Fig. \ref{Histogramas}, we present the constructed histograms, elucidating the distributions of velocity and velocity dispersion, each weighted by luminosity. The top left histogram in Fig. \ref{Histogramas} delineates the luminosity-weighted velocity histogram, highlighting the predominant velocities within the ionized gas. A noteworthy observation from this histogram is the resemblance in shape between the single-component fit and the principal double-component fit, with both showcasing peaks at around -5 km/s, a velocity that aligns with the brightest part of the MB of IZw18, the North knot. These components also presents a secondary peak at around 30 km/s which correspond with the South knot. However, the secondary double-component fit displays a more dispersed pattern in velocity, extending the breadth of the observed velocities.

Conversely, the top right histogram in Fig. \ref{Histogramas} illustrates the luminosity-weighted velocity dispersion histogram, shedding light on the prevailing velocity dispersions within the galaxy. Similar to the velocity histogram, the single-component fit in the velocity dispersion histogram bears a resemblance to the principal double-component fit in shape. Nevertheless, there are stark differences in the values they reach. The principal double-component fit achieves lower velocity dispersions, with its median situated around 15 km/s, whereas the median of the single-component fit is approximately 20 km/s. The secondary double-component fit, on the other hand, reveals markedly higher velocity dispersion, with a median of 50 km/s and being much more spread that the other components. 

The middle panels in Fig. \ref{Histogramas} show the histograms of the Halo
NE of IZw18, offering insights into the luminosity-weighted distributions of
velocity and velocity dispersion. The middle left histogram in Fig. \ref{Histogramas} distinctly outlines the luminosity-weighted velocity histogram, emphasizing the prevailing velocities within the ionized gas of this region. On the other hand, the middle right histogram in Fig. \ref{Histogramas} portrays the luminosity-weighted velocity dispersion histogram, revealing the dominant velocity dispersions within the Halo NE of the galaxy.

Upon close examination of these histograms, we observe stark kinematic distinctions between the principal and secondary components of the double-component fit. The principal component is distinctly shifted to bluer, or lower, velocities, exhibiting a median at around -30 km/s and presenting a relatively low velocity dispersion with a median around 13 km/s. In contrast, the secondary component is skewed towards redder velocities, with a median at around 3 km/s, and manifests a higher velocity dispersion, characterized by a median around 32 km/s.

This pronounced separation in the kinematics of the different components is corroborated by the velocity and velocity dispersion maps of the NE Halo (see Figs. \ref{Velocidad_2_Gauss_Principal},\ref{Velocidad_2_Gauss_Secundaria},\ref{Dispersion_2_Gauss_Principal} and \ref{Dispersion_2_Gauss_Secundaria}), which vividly depict the disparate kinematic patterns. Interestingly, the single-component fit occupies a middle ground between these two configurations, serving as a poignant illustration of the limitations inherent in single-component fits in capturing the full spectrum of kinematic information. Indeed, even in the regions of the galaxy that are relatively kinematically quiescent (i.e. those presenting lower velocity differences and reduced velocity dispersion) significant kinematic diversity is observable, underscoring the necessity of employing nuanced analytical approaches to unravel the multifaceted kinematic landscape of the galaxy.

%


Similarly, for the Halo SW of IZw18, our focus shifts to the bottom histograms in Fig. \ref{Histogramas}. These figures unveil the developed histograms, demonstrating the distributions of velocity and velocity dispersion, each weighted by luminosity, within this segment of the galaxy. Bottom left histogram in Fig. \ref{Histogramas} offers an in-depth perspective on the luminosity-weighted velocity histogram, highlighting the prominent velocities within the ionized gas of the Halo SW. Conversely, bottom right histogram in Fig. \ref{Histogramas} reveals insights into the luminosity-weighted velocity dispersion histogram, illustrating the predominant velocity dispersions in this region.

A salient observation here is the comparable luminosities between the principal and secondary components of the double-component fit, rendering the distinction between a principal and a secondary component somewhat ambiguous in this instance. In terms of velocity, the principal component is bifurcated into two regions, around -15 km/s and 45 km/s, with the latter  leaning towards the redder, or higher velocities. The secondary component has its median also around -15 km/s but extends to bluer, or lower velocities, distinguishing it as the bluer component this time. The single-component fit situates itself between these two components, without reaching the extreme velocities exhibited by them.

In relation to velocity dispersion, the observations here parallel those in other regions of the galaxy, with the principal component exhibiting lower and the secondary component displaying higher and more varied velocity dispersion. However, it is noteworthy that the shape of the single component in this region aligns more with the secondary component than with the principal one.

An additional noteworthy point is the heightened kinematic excitement in the Halo SW compared to the Halo NE. This added layer of kinetic complexity likely intensifies the intricacy of the system, making the interpretations more challenging. This nuance underscores the importance of meticulous analysis in decoding the diverse and intricate kinematic patterns inherent in different regions of the galaxy.

%

This comprehensive approach, encompassing different regions of the galaxy, furnishes a holistic understanding of the disparate kinematic behaviors and patterns pervading throughout the various segments of IZw18.

\section{Integrated spectra}
\label{Section:Integrated spectra}

We also took advantage of our IFS data to produce the 1D spectra of selected galaxy regions. These integrated spectra allow us to study each galactic region (MB knots, Halo NE and Halo SW) as a whole, while at the same time improving the signal to noise ratio (S/N).

\begin{figure*}[h!]
\centering
\resizebox{\hsize}{!}{\includegraphics{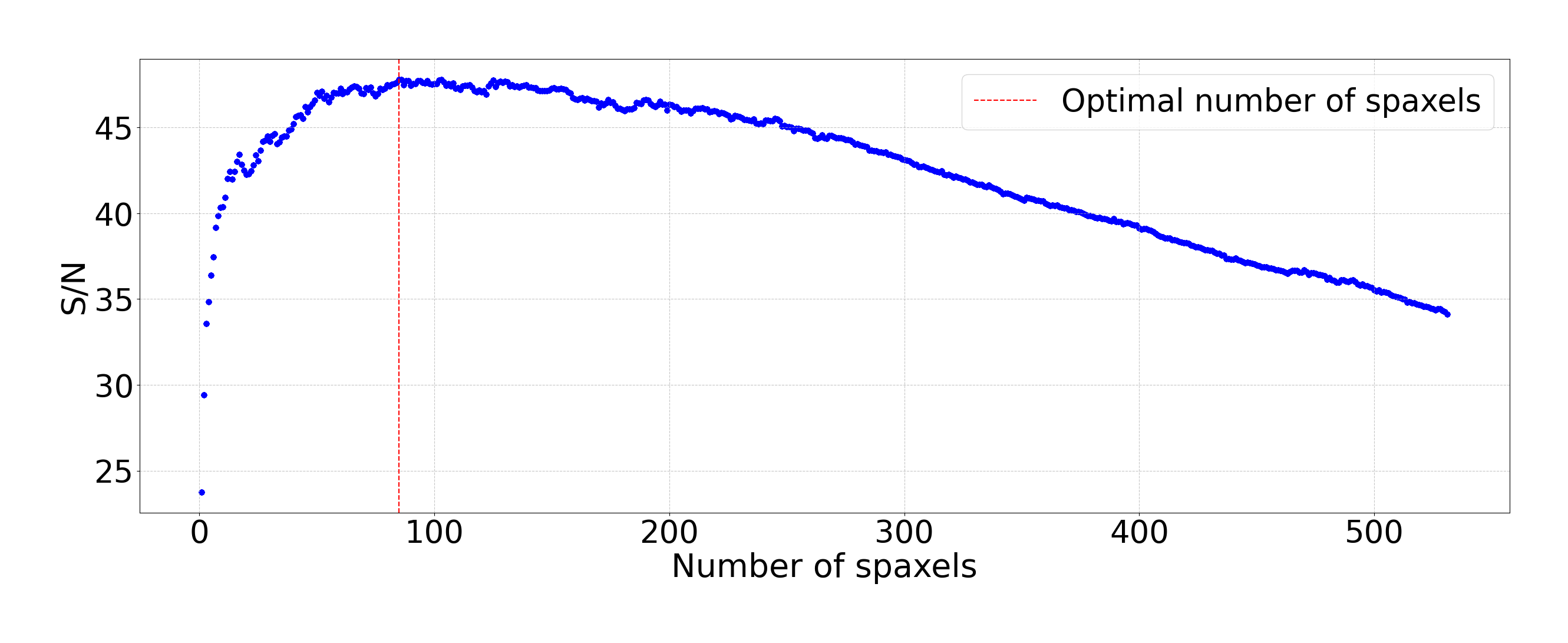}}
\caption{S/N vs number of spaxels in the MB of IZw18}
\label{SN_vs_spaxels}
\end{figure*}

\begin{figure*}[h!]
\centering
\resizebox{\hsize}{!}{\includegraphics{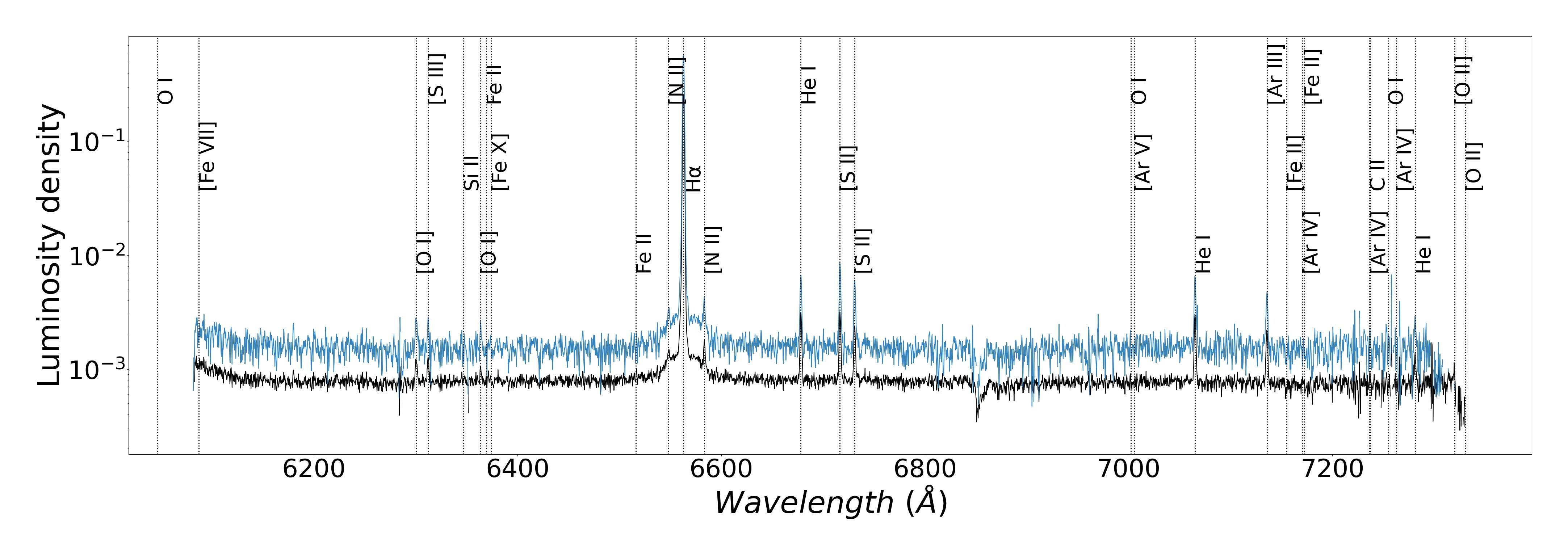}}
\caption{Integrated spectrum of the MB. In black it is represented the spectrum which optimizes S/N, in blue it is represented the integrated spectrum where all spaxels were used.}
\label{MB_espectro_completo_comparacion}
\end{figure*}

While the integrated spectra (for each region) could be constructed using all available spaxels, a more discerning approach is employed to optimize the quality of the spectra obtained. Rather than indiscriminately incorporating every spaxel, we strategically add one spaxel at a time, starting with the most luminous ones, to strike a balance between enhancing signal strength and minimizing noise interference. This method allows us to identify a point where the S/N is optimized. We halt the addition of spaxels at this point to avoid diluting the quality of the spectra with excessive noise from less luminous spaxels. In essence, this meticulous approach ensures the resultant integrated spectra are both robust and representative of the true characteristics of the selected galactic regions. As a way to demonstrate this approach in Fig \ref{SN_vs_spaxels} is represented the S/N vs the number of spaxels in the MB of IZw18. The signal (S) is the H$\alpha$ luminosity, meanwhile the noise (N) is the standard deviation of the whole continuum times the Full Width at Zero Intensity (FWZI). This allows us to compare the area of the profile with an area defined by the noise, which makes S/N to be dimensionless. We can see that the maximum S/N is reached using the 85 brightest spaxels in this case. Furthermore, in Fig. \ref{MB_espectro_completo_comparacion} we represent the optimal spectrum of the IZw18 MB in black along the integrated spectra retrieved from taking in to account all spaxels. We can clearly see that the optimized spectrum shows a lower lever of noise than the other one.

In the integrated spectra of the MB of IZw18, we observe a rich variety of spectral lines beyond the H$\alpha$ emission. Notably, the spectrum reveals the presence of the [NII]$\lambda\lambda$6548,6584
 doublet and the [SII]$\lambda\lambda$6716,6731 doublet, as well as [OI]$\lambda$6300, [SIII]$\lambda$6312, recombination lines from HeI, and [ArIII]$\lambda$7135. These additional spectral features are evident in Fig. \ref{MB_espectro_completo_comparacion}. In contrast, the spectral analysis of both Halos of IZw18 presents a markedly different scenario. Here, we observe only the $H\alpha$ line, with no discernible presence of other spectral lines. This is attributed to the faintness of the emission in these regions. 

\begin{figure*}[h!]
\centering
\resizebox{\hsize}{!}{\includegraphics{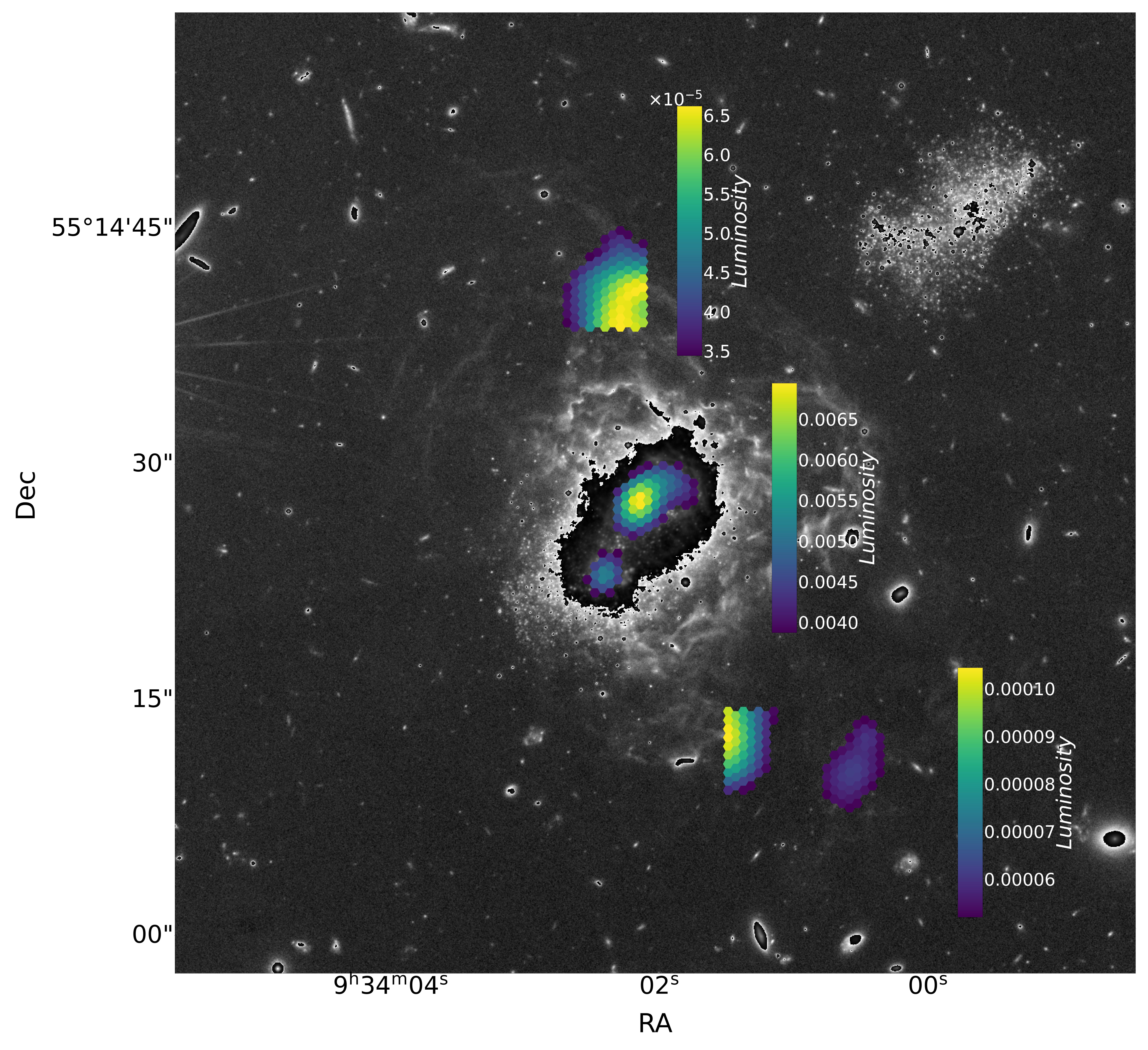}}
\caption{Luminosity map of the single-component fit. The spaxels represented here are the ones used for making the integrated spectra in each region.}
\label{Luminosidad_integrado}
\end{figure*}

To provide a visual representation of our selective approach in constructing integrated spectra, in Fig. \ref{Luminosidad_integrado} we present the H$\alpha$ maps of the observed galactic regions showing only the spaxels that were used to create the integrated spectra . Each subfigure within this composite figure represents only the spaxels that were utilized to make the optimal integrated spectra. As we can see in this image the MB is divided in two disconnected regions, the North knot and the South knot. In this way, we have done one integrated spectrum for each knot in order to distinguish their kinematical behaviour.
We can compare this image with the one that shows all spaxels (see Fig. \ref{Luminosidad_1_Gauss}). Noticeable, only a small area comprising the brightest spaxels of each region is used in order to construct the integrated spectra.

\begin{figure*}[h!]
\centering
\resizebox{\hsize}{!}{\includegraphics{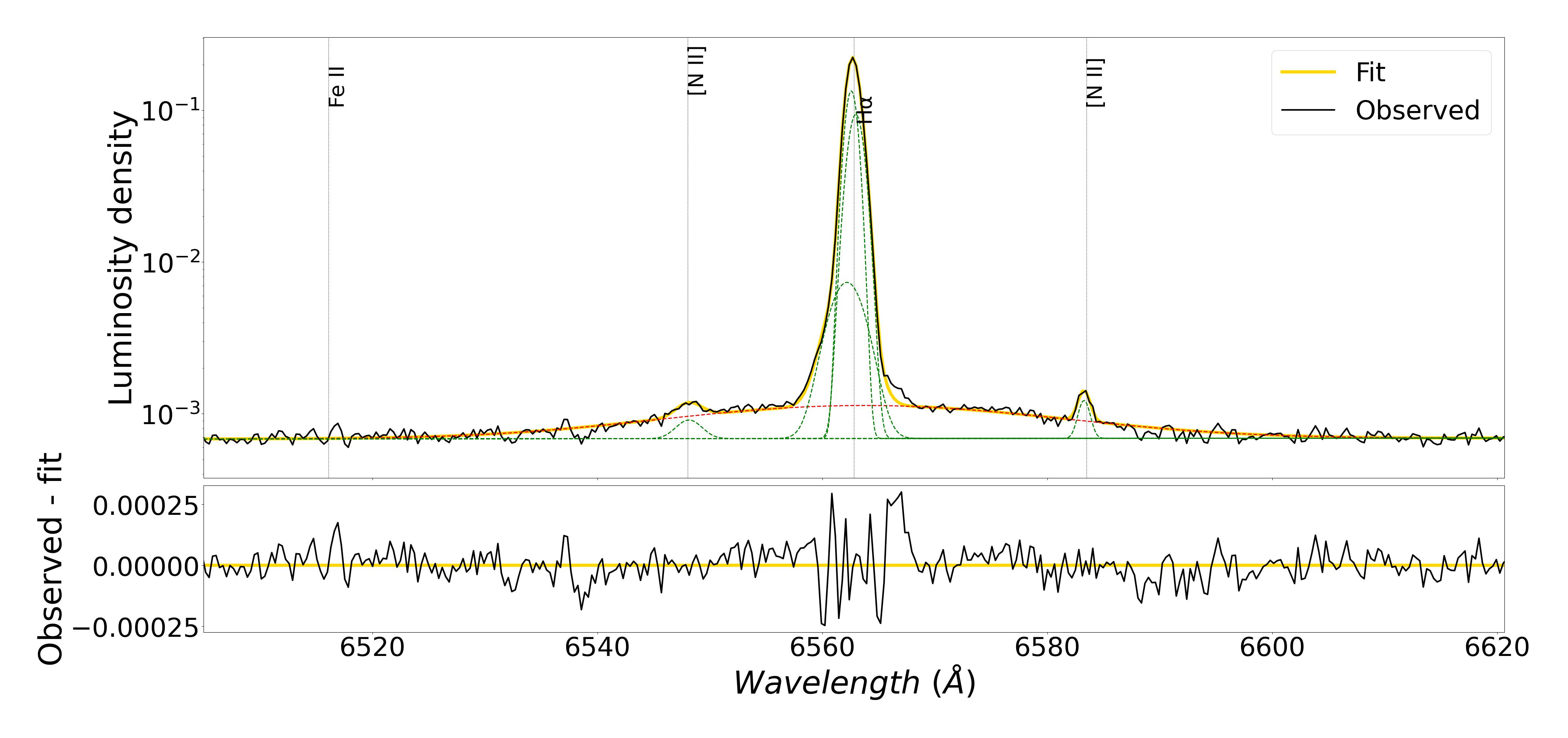}}
\caption{Zoom in to the H$\alpha$ line in the integrated spectrum of the North knot of the MB.
In black it's represented the observed spectrum and in yellow the multi-gaussian fit. In green we can see the continuum plus the individual gaussians. In red it is represented the gaussian with high standard deviation ($\sigma$>10 $\AA$).
This same approach is used in the following representation of the integrated spectra.}
\label{MB_espectro_optimaSN_Nknot}
\end{figure*}

\begin{figure*}[h!]
\centering
\resizebox{\hsize}{!}{\includegraphics{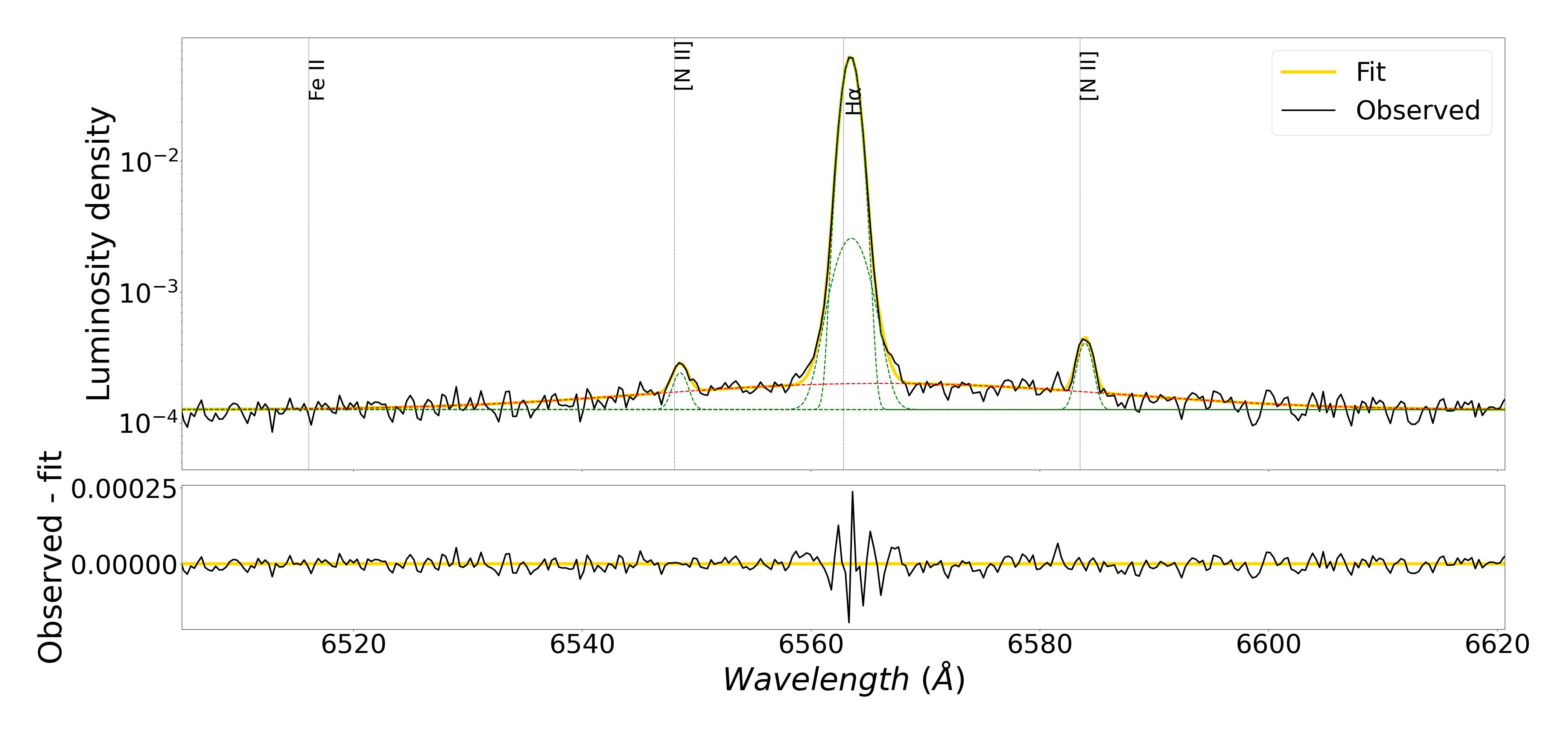}}
\caption{Zoom in to the H$\alpha$ line in the integrated spectrum of the South knot of the MB.}
\label{MB_espectro_optimaSN_Sknot}
\end{figure*}

\begin{figure*}[h!]
\centering
\resizebox{\hsize}{!}{\includegraphics{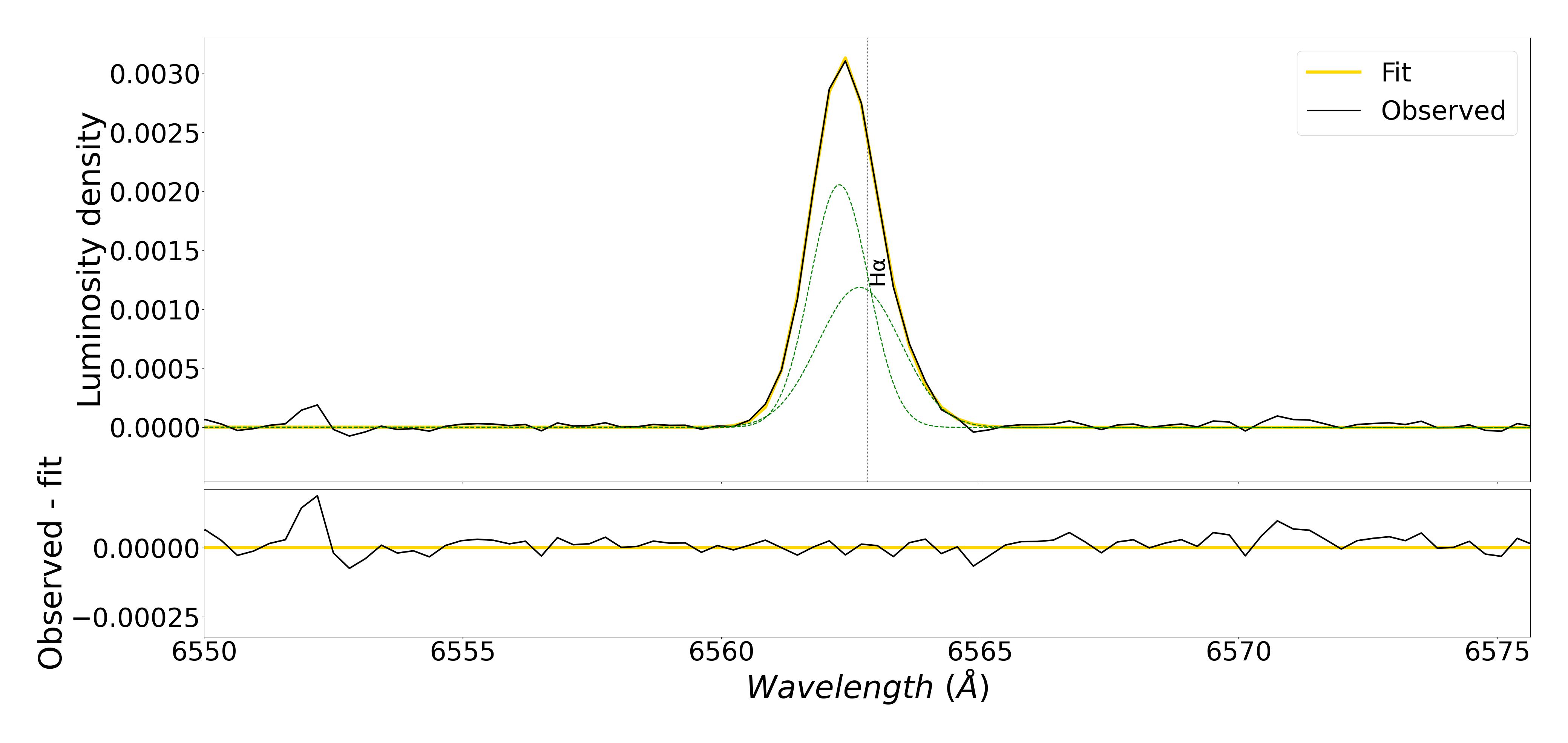}}
\caption{Zoom in to the H$\alpha$ line in the integrated spectrum of the Halo NE.}
\label{HaloNE_espectro_optimaSN}
\end{figure*}

\begin{figure*}[h!]
\centering
\resizebox{\hsize}{!}{\includegraphics{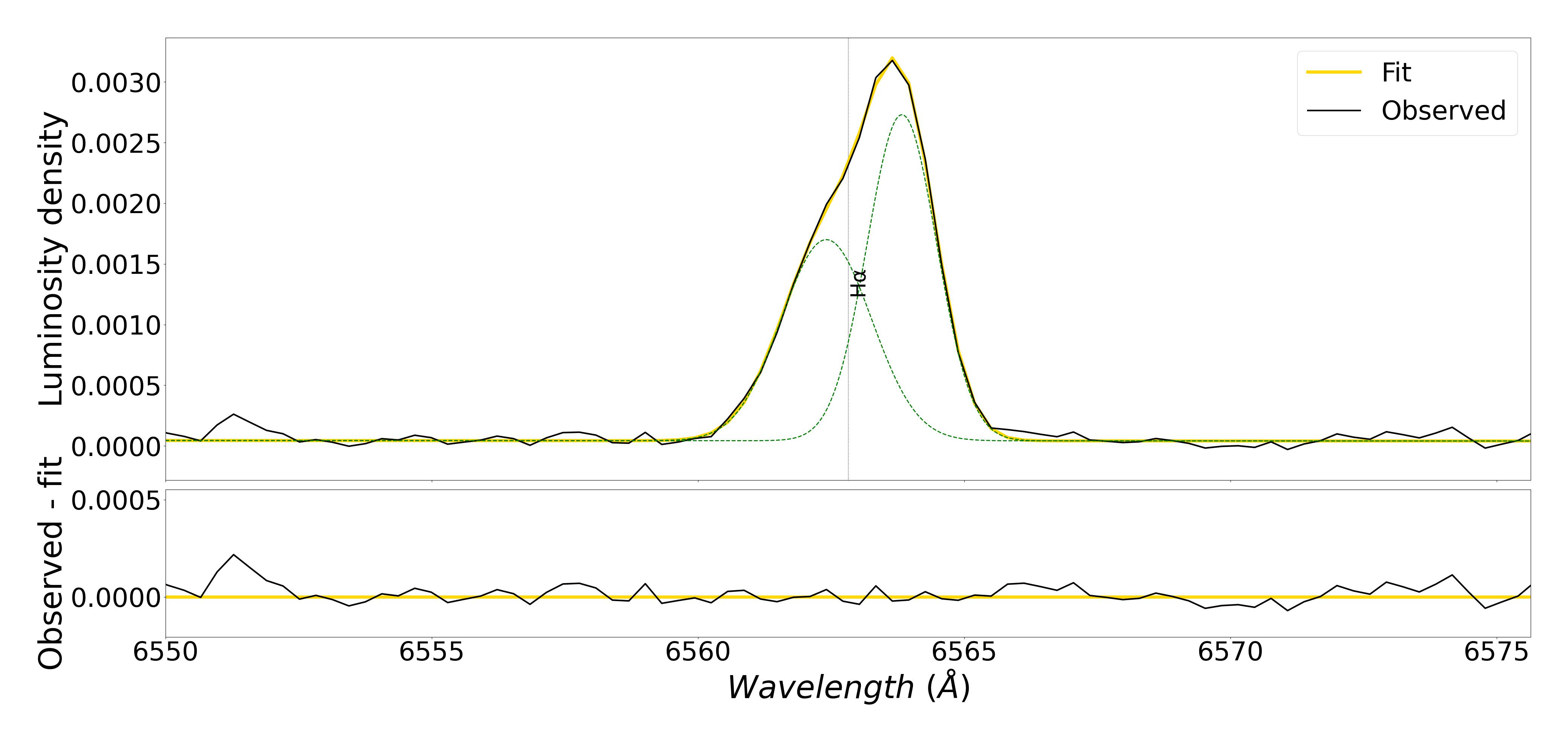}}
\caption{Zoom in to the H$\alpha$ line in the integrated spectrum of the Halo SW.}
\label{HaloSW_espectro_optimaSN}
\end{figure*}

To delve deeper into the kinematical properties of each galactic region, we present a detailed view of the H$\alpha$ line of the integrated spectra in Figs. \ref{MB_espectro_optimaSN_Nknot}, \ref{MB_espectro_optimaSN_Sknot}, \ref{HaloNE_espectro_optimaSN} and \ref{HaloSW_espectro_optimaSN}. In each region, a multi-Gaussian fit is performed on the H$\alpha$ line to extract the kinematical properties that represent these regions as a whole, utilizing the optimal S/N provided by our data. 
The AIC is employed to determine the optimal number of Gaussians for the fit, ensuring a balance between the goodness of fit and the complexity of the model. This approach allows us to accurately model the H$\alpha$ profile of each region and gain insights into the underlying kinematical structures with enhanced reliability.
To quantify the uncertainties associated with the Gaussians in the fit, we employed the bootstrapping method \citep{efron1985bootstrap}. In table \ref{table:kinematic_values} it is shown the luminosity percentage, velocity, velocity dispersion and FWZI (along with their corresponding errors) of the components that fit each region.


In examining the H$\alpha$ profile of the north (south) knot of the MB  (see Figs. \ref{MB_espectro_optimaSN_Nknot} and \ref{MB_espectro_optimaSN_Sknot}), the AIC suggests a fit with four (three) Gaussian components for the H$\alpha$ line, alongside two additional Gaussians for the [NII] lines.  The four (three) Gaussians comprise two (one) principal components, which account for $\sim$ 90$\%$ (in both cases) of the total luminosity and exhibit low velocity dispersion (from 6 km/s to 21 km/s), and two additional components, contributing around 7\% and 4\% to the luminosity (in both cases), characterized by velocity dispersions of 65 km/s (56 km/s) and an exceedingly broad velocity dispersion of 730 km/s (840 km/s), respectively (see Table \ref{table:kinematic_values}). Furthermore, the FWZI for the North and South knots is measured to be 3000 km/s and 2600 km/s, respectively, underscoring the extensive breadth of the H$\alpha$ profile, particularly in the context of the very broad components. 

The emergence of these broad and very broad components introduces a complex layer to our understanding of the kinematical structure of the galaxy. Notably, the very broad components not only are found shifted to the red in comparison to the narrow components by at least $\sim$ 40 km/s, but they also exhibit a velocity spatial gradient that is almost three times that of the narrow components. Specifically, the velocity difference between the south and North knot for the very broad components is approximately 110 km/s, which, after adjusting for the 40 km/s attributed to the rotation of the galaxy, results in a net difference of 70 km/s. This finding indicates that the very broad components are not only redshifted relative to the narrow ones but also suggest that the dynamics of the very broad components are significantly distinct, highlighting that a fraction of the ionized gas in the MB of IZw18 possesses remarkably high kinetic energy, potentially due to dynamic processes within the galaxy.

The likely physical mechanisms that could impart such a significant input of kinetic energy into the gas include supernovae explosions and/or black hole accretion. Supernovae, with their colossal energy output, can create shockwaves that propagate through the interstellar medium, thereby inducing turbulent motion in the gas. Similarly, if there is a black hole present in IZw18, its accretion activity and associated energetic outputs (such as jets or outflows) could also serve as a potent source of kinetic energy, stirring the surrounding gas into a frenzied, disordered state. Even broader components (FWHM>$10^4$ km/s) are seen in a similar (very metal-poor dwarf) galaxy called SBS 0335-052E 
 in \cite{hatano2023active}, where they argued about the possibility of an AGN being the source behind this broad emission. The discernment of this very broad component not only challenges our previous kinematical analyses but also beckons a deeper exploration into the internal dynamics and potential energy sources within IZw18, offering a rich avenue for further study and discussion.

The absence of discernible indicators of these broad components in earlier stages and the fact that so many components are found in the integrated spectra of the MB of IZw18 indicate that the dual-component fit in the previous 2D-kinematical analysis could be conflating these (up to four) components, especially in the secondary one. This revelation necessitates a reevaluation of the kinematical structure and the potential interplay between these components. 
Currently, our 2D-kinematical study extends up to the dual-component fit, emphasizing the two predominant (brightest) components discerned in the fit of the integrated spectra. The newfound complexity and multiplicity of components in the integrated spectral data invite further, more granular exploration into the kinematical properties of IZw18.

In the context of the Halo NE of IZw18, the integrated spectrum discernibly align with a two-component fit (see Fig. \ref{HaloNE_espectro_optimaSN}). Our comprehensive 2D-kinematical examination of this region corroborates this dual-component structure, revealing distinct kinematics in both the maps representations (see Figs. \ref{Velocidad_2_Gauss_Principal}, \ref{Velocidad_2_Gauss_Secundaria}, \ref{Dispersion_2_Gauss_Principal} and \ref{Dispersion_2_Gauss_Secundaria}). 
The two components manifest as a brighter, blue-shifted element with lower velocity dispersion and a dimmer, red-shifted counterpart exhibiting higher velocity dispersion.
This dual-structure is consistently evident in the integrated spectrum fit (see Fig. \ref{HaloNE_espectro_optimaSN}), where both components are distinctly recognizable. The Halo NE stands out as a region of relative kinematic simplicity, featuring two discernible components that can be readily attributed to distinct ionized gases exhibiting differing kinematics. This scenario exemplifies a case where our mathematical approach of Gaussian fitting converges with direct physical interpretations, offering insights into the movements of the ionized gas within this region.

Observations of the Halo SW of IZw18 reveal a consistent two-component fit within the integrated spectra for this region (see Fig. \ref{HaloSW_espectro_optimaSN}), mirroring the dual-structure found in the Halo NE. However, the components here exhibit an inverse configuration compared to those in the Halo NE. Specifically, the brighter component is red-shifted and exhibits a wider spread, contrasting with the narrower, blue-shifted dimmer component. The components in this region tend to have higher velocity dispersion, indicative of a broader kinetic range, reflecting the inherent kinematic complexity of the Halo SW already seen in \ref{subsection: Luminosity, velocity and velocity dispersion maps}.


\begin{table*}[h]
\centering
\caption{H$\alpha$ fit kinematical values.}
\label{table:kinematic_values}
\begin{tabular}{cccccc} 
\hline \hline
\textbf{Galactic region} & \textbf{Component} & \textbf{Luminosity}         & \textbf{Velocity}     & \textbf{Velocity dispersion ($\sigma$)} &\textbf{FWZI}\\ 
                &           &       $\%$         & (km/s)       & (km/s)                &(km/s)\\
 \textbf{(1)}& \textbf{(2)}& \textbf{(3)}& \textbf{(4)}& \textbf{(5)}&\textbf{(6)}\\ 
\hline       
MB North knot& Principal & 0.5±0.2& -11.0±0.8& 6±2&3000±200\\
                & Secondary &  0.4±0.2&  10±8& 21±3&\\
                & Broad      &  0.064±0.004& -29±3& 65±2&\\
 & Very broad& 0.046±0.002& 40±40&730±30&\\
 \hline   
 MB South knot& Principal& 0.896±0.006& 28.94±0.02& 17.0±0.08&2600±300\\
 & Broad& 0.074±0.004& 32.0±0.7& 56±2&\\
 & Very broad& 0.031±0.003& 150±80& 840±90&\\
\hline
Halo NE& Principal & 0.7±0.3& -23±3& 12±4&210±40\\
                & Secondary &  0.3±0.3&  0±30& 20±20&\\
\hline
Halo SW& Principal & 0.6±0.2& 46±4& 18±5&270±30\\
                & Secondary &  0.4±0.2& -20±20& 30±8&\\
\hline
\end{tabular}

\begin{tablenotes}
      \small
      \item Column (1): Name of the galactic region. Column (2): Name given to each gaussian component. Column (3): Normalized luminosity corresponding to each component. Column (4): Velocity corresponding to each component. Column(5): Velocity dispersion corresponding to each component. Column(6): FWZI of the line profile.
\end{tablenotes}

\end{table*}

\section{Very broad component}

Based on the previous discovery of the very broad component present in the knots of the MB and the striking different velocities of this very broad component which respect the narrow ones we have made a similar approach than the one used in Section \ref{subsection: Luminosity, velocity and velocity dispersion maps}. This approach enabled us to generate luminosity, velocity, and velocity dispersion maps specifically for this very broad component. 

Nevertheless, as we have noted before in Fig. \ref{doble_pico} the very broad component can not be seen in individual spaxels. One way to solve this is to make a harder Gaussian spatial smoothing to the spaxels, in this way at the cost of further loosing of spatial resolution we are able to reach a point in which we have enough signal to distinguish this dim component from the continuum. This time, the Gaussian kernel used has a FWHM of 2.5 '' which enlarges the area of the FWHM of the Gaussian kernel 3 times the previous one used.

Furthermore, in order to properly fit the very broad component we have to get rid of the more luminous narrow H$\alpha$ emission and the [NII] lines. For this we take away the points in the spectra that distance less than 5 $\AA$ (2.5 $\AA$) from the center of the H$\alpha$ line ([NII] lines).

\begin{figure*}[h!]
\centering
\resizebox{\hsize}{!}{\includegraphics{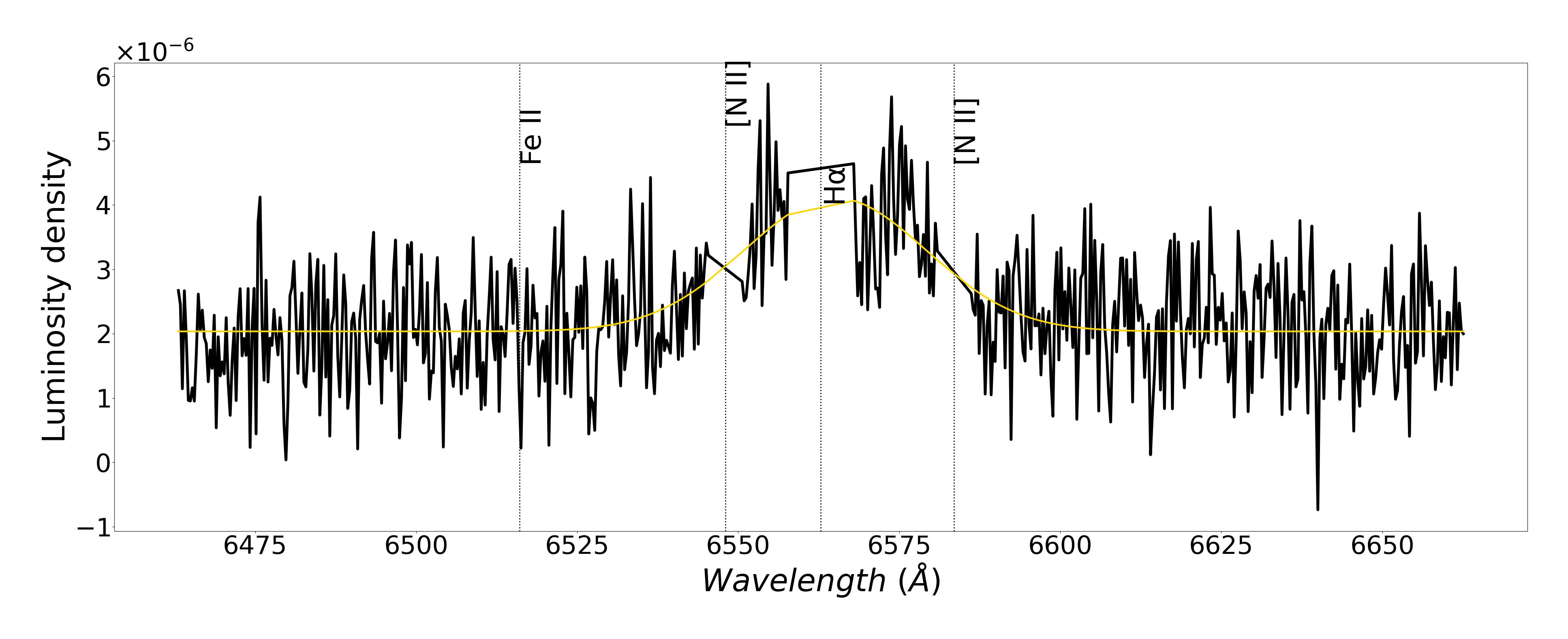}}
\caption{One-component fit of the defined broad H$\alpha$ component. The profile comes from the same spaxel as used in Fig. \ref{doble_pico}. The profile is depicted in black, meanwhile the one-component fit is represented in yellow.}
\label{Espectro_ancha}
\end{figure*}

Now we make a one-component Gaussian fit to the profiles that appears after all this procedure. The fit corresponding to the same spaxel as Fig. \ref{doble_pico} is represented in Fig. \ref{Espectro_ancha}. Here we can see that thanks to the enhanced smoothing of the new kernel the very broad component is clearly visible and can be well fitted.

The same criteria for excluding spaxels  as the one applied in Section \ref{Section:2D Kinematical analysis} is done here. Particularly, in this case the S/N ratio below 3 criterion is the one that takes away most of the spaxels.

To effectively compare these new results with the previously done one-component fit in Section \ref{Section:2D Kinematical analysis}, we have reapplied the one-component fitting procedure to the entire H$\alpha$ line, employing the new Gaussian kernel. This approach reaffirms the previously presented results, but with enhanced smoothing.

\begin{figure*}[h!]
\centering
\resizebox{10cm}{!}{\includegraphics{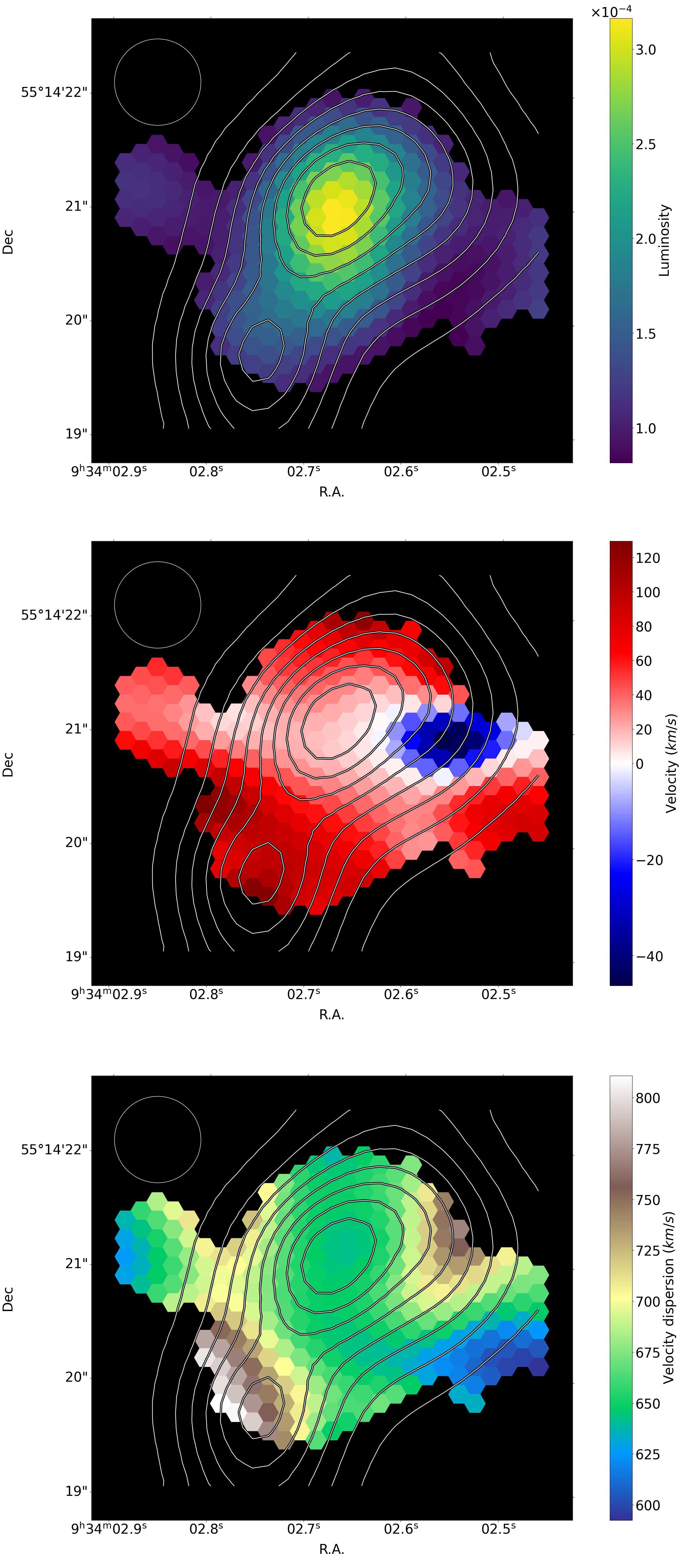}}
\caption{Three maps corresponding to the very broad component. Top map correspond to the luminosity, middle map correspond to the velocity and bottom map represents the velocity dispersion. In all maps we can see overplotted the contours of the total H$\alpha$ luminosity and a circle representing the FWHM of the seeing.}
\label{composite}
\end{figure*}

\begin{figure*}[h!]
\centering
\resizebox{15.5cm}{!}{\includegraphics{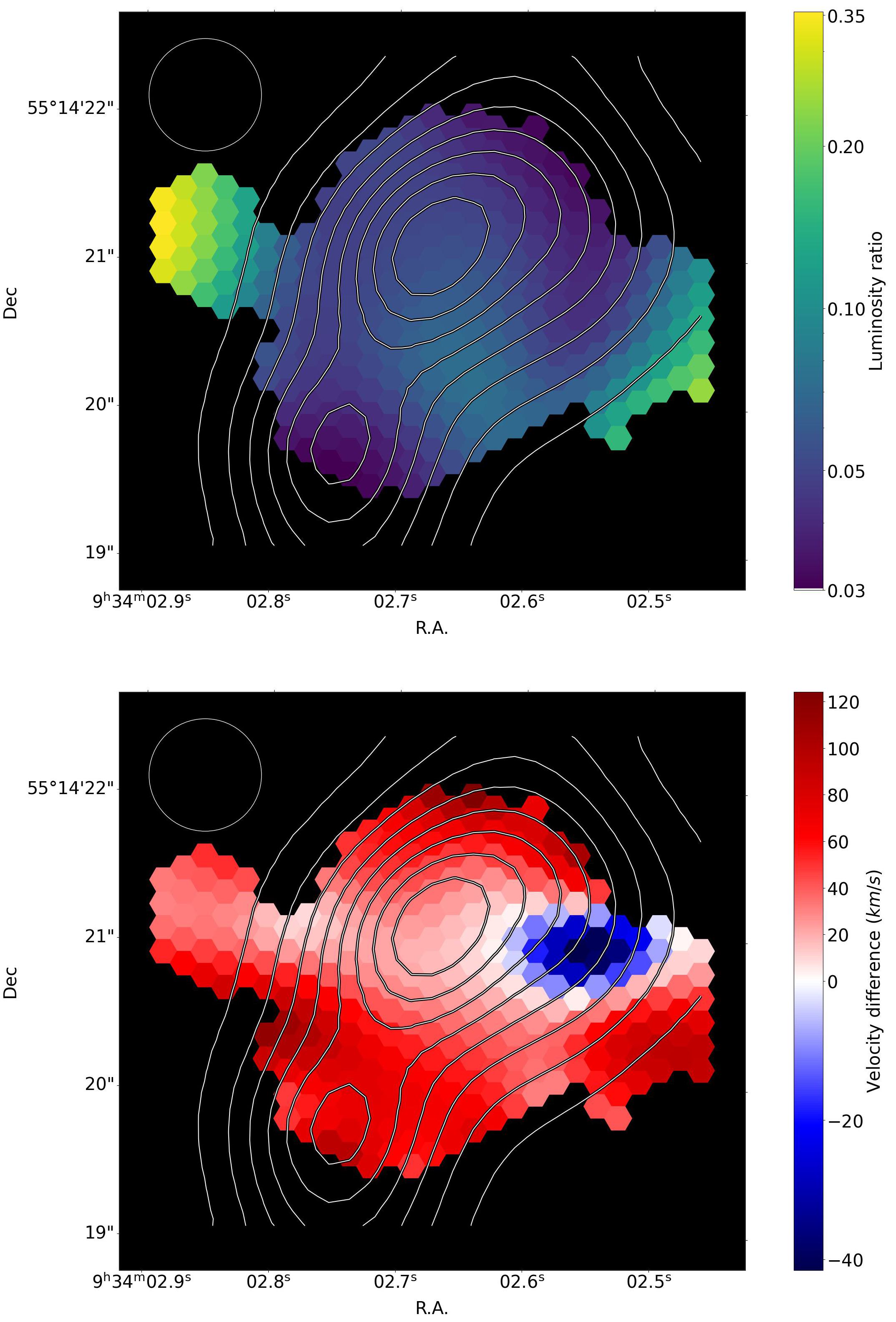}}
\caption{Top map correspond to the luminosity ratio: Very broad component luminosity / Full H$\alpha$ luminosity. Bottom map represents the velocity difference: Very broad component velocity - Full H$\alpha$ velocity. In all maps we can see overplotted the contours of the total H$\alpha$ luminosity and a circle representing the FWHM of the seeing.}
\label{composite_2}
\end{figure*}

The maps of the very broad component are represented in Fig. \ref{composite} and luminosity ratio (Very broad component luminosity / Full H$\alpha$ luminosity) and velocity difference (Very broad component velocity - Full H$\alpha$ velocity) maps are in Fig. \ref{composite_2}. 

The maximum of emission of the very broad component is slightly to the south of the maximum of the total H$\alpha$ emission (see top map in Fig. \ref{composite}). Nevertheless, the maximum luminosity ratio is in the eastern and western part of the MB (see top map in Fig. \ref{composite_2}), where we reach up to 35 $\%$ of the emission in the form of this very broad component. These regions are actually the same that presents a velocity dispersion of around 200 km/s in the secondary component of the dual-fitting (see Fig. \ref{Dispersion_2_Gauss_Secundaria}). In this way, as we mentioned before, this secondary component is fitting several components of the H$\alpha$ line profile seen in the integrated spectra of the knots (see Figs. \ref{MB_espectro_optimaSN_Nknot} and \ref{MB_espectro_optimaSN_Sknot}), making it difficult to interpret.
In the South knot we do not see any secondary maximum in luminosity regarding the very broad component, as it is seen in the luminosity of the whole H$\alpha$ line (see top map in Fig. \ref{composite}). This is reflected in the luminosity ratio map where in this region its value is reach a minimum of 3$\%$. The south part of the North knot along with the previously mentioned eastern and western regions of the MB present the maximum luminosity ratio, indicating possible zones dominated by outflows.
The wide extended spatial region that shows very broad component emission indicates that the reason behind this high kinematic input in the gas should not be local, as we could expect from localized supernovae or AGN in the North knot.

Furthermore, looking at the velocity and velocity dispersion maps (see middel and bottom maps in Fig. \ref{composite}) we recall the tremendous kinematic excitement of this gas. The velocity has a range of more than 160 km/s which is more than 3 times the velocity range presented in the one-component fit of the MB. The velocity dispersion does not present much spatial variation ranging from 600 km/s to more than 800 km/s in the South knot. The velocity difference between the very broad component and the whole H$\alpha$ line (see bottom map in Fig. \ref{composite_2}) shows that the very broad component is mostly shifted to the red with respect the narrow component. Only in the west of the North knot this difference in velocity goes to zero and even switches to the blue. 
This high velocity difference between the narrow and the very broad component indicates that these gases maybe occupy a different region along the line of sight (being the very broad component gas likely behind the narrow component gas).
A possible explanation of this behaviour is the presence of high density gas near the origin of kinematic input. This high density gas could act like a wall, that to some extent reflects back the momentum of the gas itself, producing the observed change in the velocity of the low density gas in an opposite direction. In this way, in the red velocities of the bottom map in Fig. \ref{composite_2} this theoretical wall would be in the front part of the galaxy which respect the line of sight (closer to us). On the other hand, the blue velocities in this same map indicates that this high density gas is in the back of the galaxy with respect to the line of sight (far away from us). 
This kinematic interaction of different gases within the ISM is reminiscent of the turbulent mixing layer (TML) mechanism, where emission-line broadening occurs at the interface of cold disc gas and warmer outflow gas, as discussed by \cite{hogarth2020chemodynamics}. The observed shifts in velocities between broad and narrow components have been seen in other galaxies like GP J 0820 \citep{bosch2019integral}. This might be explained by this reflection of momentum along the line of sight. In fact, this could be a result of the high-density gas acting as a barrier, influencing the dynamics of the gas flow and leading to the velocity shifts seen in the spectral lines.

\section{Conclusions}
\label{Section:Conclusions}

This study presents a comprehensive analysis of the kinematical behaviour of the ionized gas in IZw18, a galaxy known for its extremely low metallicity and proximity. Our research delves into the kinematic complexities of IZw18, exploring its structure and dynamics through an in-depth examination of the H$\alpha$ line profiles across different regions. By applying both single and double Gaussian component fittings to the spaxel data, we successfully generated luminosity, velocity, and velocity dispersion maps for each Gaussian component in the MB of the galaxy and two distinct regions of the Halo (NE and SW Halo). Additionally, our methodology included creating integrated spectra for select regions of the galaxy, focusing on maximizing the S/N ratio. Finally, we extended our analysis by constructing luminosity, velocity, and velocity dispersion maps of the very broad component (discovered in the integrated spectra), employing a larger Gaussian kernel to enhance the signal, thereby allowing for a more detailed observation of this elusive component.

In the MB, our analysis reveals a notable rotational pattern connecting the north and South knots. However, the secondary double-component fit uncovers a different and more complex kinematic pattern. This complexity is further highlighted in the integrated spectra of both knots, where up to four Gaussian components were required to adequately fit the H$\alpha$ profile. Notably, one of these components exhibited a FWHM of almost 2000 km/s, accounting for 4$\%$ of the total flux, suggesting the presence of a high-energy outflow. Particularly, this very broad component exhibits a significant spatial extent that not only challenges the notion of localized high-energy events, such as supernovae or AGN activity, but also questions the likelihood of bipolar outflows. Such outflows are generally expected in scenarios involving localized sources of kinetic energy. The widespread nature of this component instead suggests more global, galaxy-scale processes at work, indicating a complex interplay of dynamics that extends beyond traditional models of localized energetic events.

The Halo regions of the galaxy reflect the behavior of the shells in IZw18. The NE Halo exhibits a relatively tranquil kinematic state, with two Gaussian components in the integrated spectra showing low velocity dispersion ($\leq$ 20 km/s). These components align with the two components of the double-component fit, presenting bluer velocities than the MB. In contrast, the SW Halo is characterized by a more complex kinematical behaviour than the NE Halo. This is reflected in the presence of higher velocities and velocity dispersions and more complex kinematical patterns in the maps.

This research aligns with and expands upon the existing body of work on the kinematics of dwarf galaxies, particularly those with characteristics similar to IZw18, a blue compact dwarf galaxy. Consistent with observations in other galaxies of its class, our study confirms that IZw18 exhibits chaotic kinematics alongside a discernible disk rotation pattern. This finding suggests that while a primary rotational pattern encompassing the knots of the MB is evident, a secondary and more intricate mechanisms are also at play. These mechanisms, possibly stemming from stellar feedback, introduce additional kinetic energy into the ISM, resulting in a more chaotic and unpredictable motion of the ionized gas of the galaxy.


Furthermore, a critical aspect of our findings revolves around the significant velocity differences observed between the narrow and the very broad components of the H$\alpha$ line. This disparity hints at a distinct spatial segregation of these gases within the line of sight, with the very broad component likely residing behind the narrow component. Intriguingly, this spatial arrangement might be influenced by the presence of high-density gas near the origin of the kinematic input. This dense gas could function to some extent as a wall, reflecting back the momentum of the surrounding gas. Such a reflection would manifest as a change in the velocity of the center of mass of the gas, directed opposite to this 'wall' of dense gas. In this way, the redshifted velocities on the bottom map in Fig. \ref{composite_2} suggest that this dense gas is located towards the front of the galaxy, relative to the line of sight (closer to us). Conversely, the blueshifted velocities imply that the high-density gas lies towards the back of the galaxy (farther from our perspective). These observations provide a novel interpretation of the internal gas dynamics in IZw18, pointing to complex, multi-layered kinematic interactions within the galaxy, and challenging conventional understandings of gas motion and distribution in such environments.

The study of IZw18, with its notably low metallicity, offers a crucial window into understanding the early galaxy formation and evolution in the Universe. This metal-poor environment in IZw18 closely resembles the conditions prevalent in the first galaxies, providing a unique opportunity to observe and infer behaviors and processes that were likely common in the high-redshift Universe. Our proximity to IZw18, coupled with the high spatial resolution of our observations, has allowed us to discern its kinematic complexity in unprecedented detail. This complexity, likely a characteristic of many galaxies, becomes vividly apparent in IZw18 due to our advantageous observational position. However, the intricate nature of these kinematic patterns presents significant challenges in interpretation, underscoring the sophisticated dynamical processes at play in such early-stage galactic environments. Through IZw18, we gain insights into the nascent stages of galaxy evolution, observing firsthand the intricate dance of forces shaping galaxies at the dawn of the cosmos.

\begin{acknowledgements}
We thanks the referee for their constructive comments.
Author Antonio Arroyo Polonio acknowledges financial support from the grant CEX2021-001131-S funded by MCIN/AEI/ 10.13039/501100011033. RA acknowledges support from ANID FONDECYT Regular Grant 1202007. The authors acknoledge the plan PID2021-123417OB-I00.
\end{acknowledgements}

\bibliographystyle{aa}
\bibliography{Biblioteca.bib}

\begin{appendix}

\section{3D kinematical representation}
\label{appendix: 3D IZw18}


\begin{figure*}[h!]
\centering
\resizebox{\hsize}{!}{\includegraphics{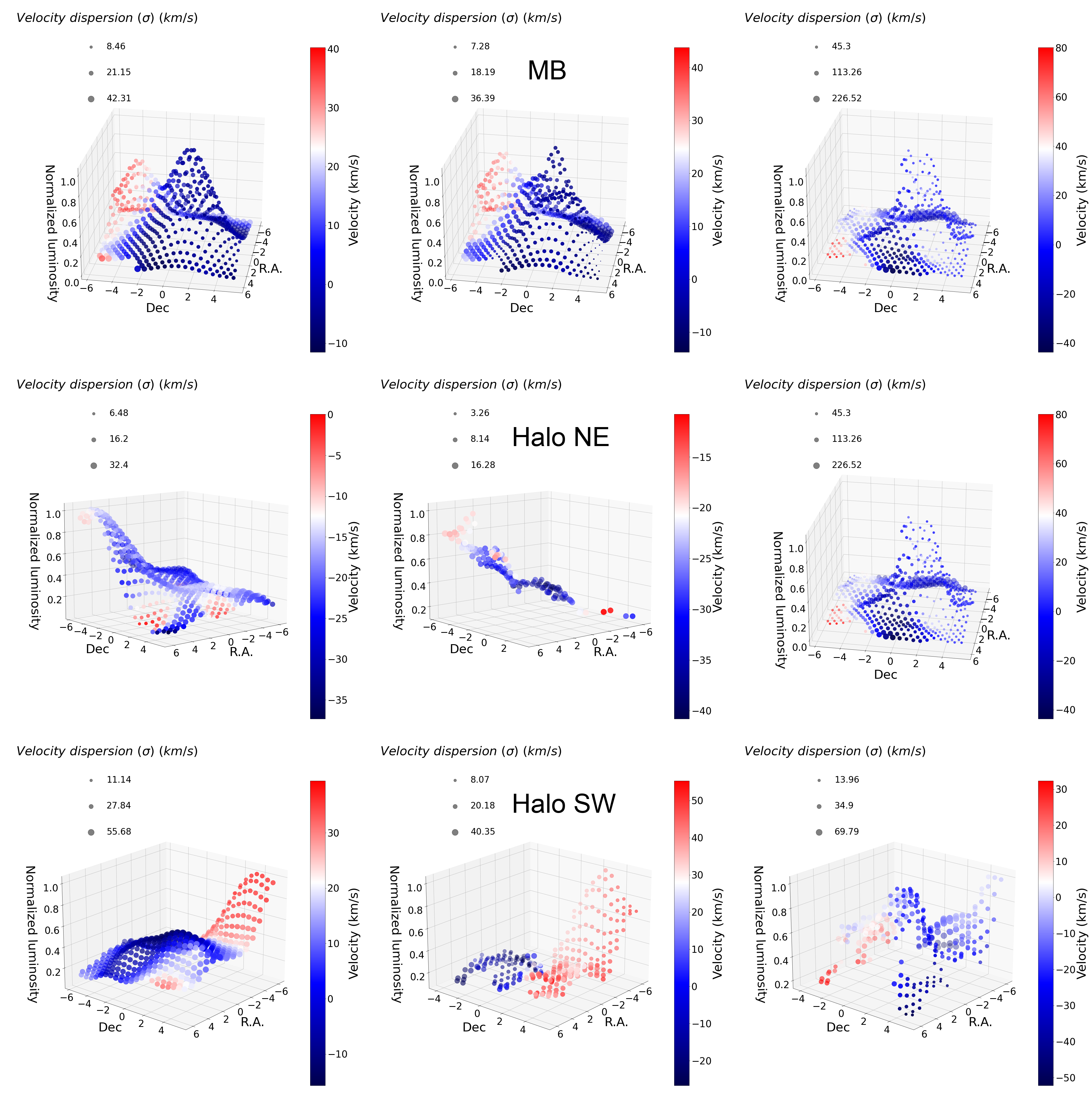}}
\caption{3D representation of the kinematical data of IZw18.
Top row corresponds to the MB, middle row to the halo NE and bottom row to the halo SW. In the left column we represent data from the single-component fit. In the middle column it is represented data from the principal double-component fit. The right column correspond to data from the secondary double-component fit.
Each 3D figure presents the following features. 
The x and y axis correspond to the position of the galaxy in ''. The z axis (vertical axis) represents the normalized luminosity. The colorbar correspond to the measured velocity and the size of each point is related to the velocity dispersion (as indicated in the top left legend in each 3D figure).}
\label{composite_3D}
\end{figure*}

\end{appendix}

\end{document}